\documentclass[onecolumn,showpacs,preprintnumbers,amsmath,amssymb,eqsecnum]{revtex4}
\usepackage{dcolumn}
\usepackage{bm}
\usepackage[dvips]{graphicx}
\usepackage{wrapfig}
\usepackage{psfrag}
\begin{document}


\def\diag{\mathop{\rm diag}\nolimits}
\def\sh{\mathop{\rm sh}\nolimits}
\def\ch{\mathop{\rm ch}\nolimits}
\def\var{\mathop{\rm var}}\def\exp{\mathop{\rm exp}\nolimits}
\def\Re{\mathop{\rm Re}\nolimits}
\def\Sp{\mathop{\rm Sp}\nolimits}
\def\kp{\mathop{\text{\ae}}\nolimits}
\def\bk{{\bf {k}}}
\def\bp{{\bf {p}}}
\def\bq{{\bf {q}}}
\def\lra{\mathop{\longrightarrow}}
\def\Const{\mathop{\rm Const}\nolimits}
\def\sh{\mathop{\rm sh}\nolimits}
\def\ch{\mathop{\rm ch}\nolimits}
\def\var{\mathop{\rm var}}
\def\mK{\mathop{{\mathfrak {K}}}\nolimits}
\def\mR{\mathop{{\mathfrak {R}}}\nolimits}
\def\mv{\mathop{{\mathfrak {v}}}\nolimits}
\def\mV{\mathop{{\mathfrak {V}}}\nolimits}
\def\mD{\mathop{{\mathfrak {D}}}\nolimits}
\def\mN{\mathop{{\mathfrak {N}}}\nolimits}
\def\mS{\mathop{{\mathfrak {S}}}\nolimits}

\newcommand\ve[1]{{\mathbf{#1}}}

\def\Re{\mbox {Re}}
\newcommand{\Z}{\mathbb{Z}}
\newcommand{\R}{\mathbb{R}}
\def\mg{\mathop{{\mathfrak {g}}}\nolimits}
\def\mK{\mathop{{\mathfrak {K}}}\nolimits}
\def\mk{\mathop{{\mathfrak {k}}}\nolimits}
\def\mR{\mathop{{\mathfrak {R}}}\nolimits}
\def\mv{\mathop{{\mathfrak {v}}}\nolimits}
\def\mV{\mathop{{\mathfrak {V}}}\nolimits}
\def\mD{\mathop{{\mathfrak {D}}}\nolimits}
\def\mN{\mathop{{\mathfrak {N}}}\nolimits}
\def\ml{\mathop{{\mathfrak {l}}}\nolimits}
\def\mf{\mathop{{\mathfrak {f}}}\nolimits}
\newcommand{\ccm}{{\cal M}}
\newcommand{\cE}{{\cal E}}
\newcommand{\cV}{{\cal V}}
\newcommand{\cI}{{\cal I}}
\newcommand{\cR}{{\cal R}}
\newcommand{\cK}{{\cal K}}
\newcommand{\cH}{{\cal H}}
\newcommand{\cW}{{\cal W}}

\def\div{\mathop{\rm div}\nolimits}
\def\br{\mathop{{\bf {r}}}\nolimits}
\def\bS{\mathop{{\bf {S}}}\nolimits}
\def\bA{\mathop{{\bf {A}}}\nolimits}
\def\bJ{\mathop{{\bf {J}}}\nolimits}
\def\bn{\mathop{{\bf {n}}}\nolimits}
\def\bg{\mathop{{\bf {g}}}\nolimits}
\def\bv{\mathop{{\bf {v}}}\nolimits}
\def\be{\mathop{{\bf {e}}}\nolimits}
\def\bp{\mathop{{\bf {p}}}\nolimits}
\def\bz{\mathop{{\bf {z}}}\nolimits}
\def\bbf{\mathop{{\bf {f}}}\nolimits}
\def\bb{\mathop{{\bf {b}}}\nolimits}
\def\ba{\mathop{{\bf {a}}}\nolimits}
\def\bx{\mathop{{\bf {x}}}\nolimits}
\def\by{\mathop{{\bf {y}}}\nolimits}
\def\br{\mathop{{\bf {r}}}\nolimits}
\def\bs{\mathop{{\bf {s}}}\nolimits}
\def\bH{\mathop{{\bf {H}}}\nolimits}
\def\bk{\mathop{{\bf {k}}}\nolimits}
\def\be{\mathop{{\bf {e}}}\nolimits}
\def\bnul{\mathop{{\bf {0}}}\nolimits}
\def\bq{{\bf {q}}}

\newcommand{\oV}{\overline{V}}
\newcommand{\vkp}{\varkappa}
\newcommand{\os}{\overline{s}}
\newcommand{\opsi}{\overline{\psi}}
\newcommand{\ov}{\overline{v}}
\newcommand{\oW}{\overline{W}}
\newcommand{\oPhi}{\overline{\Phi}}

\def\mI{\mathop{{\mathfrak {I}}}\nolimits}
\def\mA{\mathop{{\mathfrak {A}}}\nolimits}

\def\st{\mathop{\rm st}\nolimits}
\def\tr{\mathop{\rm tr}\nolimits}
\def\sign{\mathop{\rm sign}\nolimits}
\def\d{\mathop{\rm d}\nolimits}
\def\const{\mathop{\rm const}\nolimits}
\def\O{\mathop{\rm O}\nolimits}
\def\Spin{\mathop{\rm Spin}\nolimits}
\def\exp{\mathop{\rm exp}\nolimits}
\def\SU{\mathop{\rm SU}\nolimits}
\def\mU{\mathop{{\mathfrak {U}}}\nolimits}
\newcommand{\cU}{{\cal U}}
\newcommand{\cD}{{\cal D}}

\def\mI{\mathop{{\mathfrak {I}}}\nolimits}
\def\mA{\mathop{{\mathfrak {A}}}\nolimits}
\def\mU{\mathop{{\mathfrak {U}}}\nolimits}

\def\st{\mathop{\rm st}\nolimits}
\def\tr{\mathop{\rm tr}\nolimits}
\def\sign{\mathop{\rm sign}\nolimits}
\def\d{\mathop{\rm d}\nolimits}
\def\const{\mathop{\rm const}\nolimits}
\def\O{\mathop{\rm O}\nolimits}
\def\Spin{\mathop{\rm Spin}\nolimits}
\def\exp{\mathop{\rm exp}\nolimits}

\title{Maxwell equations in curved space-time:  non-vanishing magnetic field in pure electrostatic systems}

\author {N.N. Nikolaev and S.N. Vergeles\vspace*{4mm}\footnote{{e-mail:vergeles@itp.ac.ru}}}

\affiliation{Landau Institute for Theoretical Physics,
Russian Academy of Sciences,
Chernogolovka, Moscow region, 142432 Russia \linebreak
and   \linebreak
Moscow Institute of Physics and Technology, Department
of Theoretical Physics, Dolgoprudnyj, Moskow region,
141707 Russia}

\begin{abstract}

Solutions of the Maxwell equations for electrostatic systems with manifestly vanishing electric currents
in the curved space-time for stationary metrics 
are shown to exhibit a non-vanishing magnetic field of pure geometric origin. In contrast to the conventional magnetic field of the Earth
it can not be screened away by a magnetic shielding. As an example of practical significance we treat 
electrostatic systems at rest on the rotating Earth and derive the relevant geometric magnetic field. We
comment on its  impact on the ultimate precision searches of the electric dipole moments of ultracold neutrons and of protons in all electric storage rings.

\end{abstract}


\maketitle

\section{Introduction}

In this work we raise  one generic issue with Maxwell equations in curved space-time. Specifically,
we focus on the case of pure electrostatic systems in the noninertial reference frame with a globally time-independent metric tensor with $g_{0i}\neq0$.
Evidently, for an observer in a flat space-time and inertial frame the charges in motion do generate both 
electric and magnetic fields. The issue is what are the fields seen by an observer in the noninertial frame  comoving with the
electric charges, so that to this observer  the electric current is a vanishing one. Will he see pure
electric field or shall there be some trace of the noninertial motion in the form of a residual magnetic field?
From the point of view of Einstein's general relativity theory such a magnetic field if exists would be 
of geometric origin and will have a peculiar property of not to be subjected to screening by the
conventional magnetic shields.

A focus of this paper will be on the special case of noninertial motion of practical interest: magnetic fields
in electrostatic systems residing at rest in the curved space-time of the rotating body.
Our principal statement is  that the full system of Maxwell  equations entails a non-vanishing 
magnetic field despite zero currents. In terms of the
metric properties of the relevant space-time, the effect is due to the nonvanishing off-diagonal components 
of the metric tensor $g_{0i}, \ i=1,2,3$. In cases of practical interest, this geometric magnetic field is driven by the Earth rotation.

Our primary motivation was  the recent interest in the impact of the
gravity induced spin rotations in the ultimate sensitivity searches for
the electric dipole moments (EDM) of neutrons and of charged particles in the storage ring experiments \cite{Silenko2007}-\cite{Abusaif2019}.  The effect of the Earth
rotation in the neutron EDM experiments has already received much attention \cite{Baker2007},  \cite{Lamoreaux2007},
\cite{Serebrov2015}.  The impact of the
geometric magnetic field has not been discussed before. As we shall see, this field has a
strong dependence on the configuration of the static electric field.   In the  geometry  of the neutron EDM experiments, the geometric
magnetic field is a nonuniform one with the nonvanishing gradient. Such a gradient of the magnetic
fields is of particular concern in the comagnetometry which is crucial in the neutron EDM experiments (\cite{Abel2019} and references therein) . To this end, the salient feature of the geometric magnetic field is that it changes a sign when the external electric field is flipped. Consequently,
interaction of the magnetic moment of the particle with the geometric magnetic field will generate a false signal of EDM. Numerically, the false EDM is still below the sensitivity of the current neutron EDM experiments, but will be in the
ballpark of the ultimate sensitivity neutron EDM experiments aiming atvimproving the existing upper bound by  one order in magnitude and beyond \cite{Chupp2019}.

The further presentation is organized as follows. A treatment of physics in the noninertial frames demands for a consistent use of Einsten's general relativity (GR) formalism. In  Section II we invoke GR to derive the results for the geometry of the curved spfce-time and dynamics of spinning charged particles in such spaces. 
In particular, since the space-time is curved, one needs to elucidate the meaning of the  
magnetic field. Specifically, we check that the magnetic  field excerts on a charged particle a force which rotates its velocity ${\bf v}$ (defined in a certain orthonormal basis) preserving its magnitude, and induces the conventional precession of its magnetic moment. Trivial though may they sound, these constraints are important, when the nonvanishing geometrical magnetic field is present.

A subject of Section III is a derivation of the geometric magnetic field in the pure electrostatic systems at rest in the noninertial reference frames. We demonstrate that the geometric magnetic field is a salient feature of Maxwell equations in the stationary metrics with $g_{0i}\neq0$. In the simplest case of the laboratory residing on the rotating Earth, a connection between the geometric magnetic field, which is an axial vector, and the polar external electric field, contains the angular velocity of  Earth's rotation, so that the parity conservation is warranted.  

In Section IV we comment on three examples of the geometric magnetic field. The most interesting case at present is a search for the EDM of neutrons via precession of the neutron spin in the external electric field. Here the geometric magnetic field happens to have a finite gradient along the electric field and gives rise to the false EDM signal. Although tiny, this false EDM is in the ballpark of future ultimate precision neutron EDM experiments. The second example is the geometric magnetic field in all electric storage rings. Such a ring is discussed as a dedicated proton EDM machine \cite{Abusaif2019,Anastopoulos2016}. In this case, in contrast to the constant residual magnetic field of the Earth, the geometric magnetic moment has a quadrupole-like behaviour along the circumference of the storage ring. Finally, for the sake of academic completeness, we report an exact solution for the uniformly charged sphere at rest on Earth's surface. 

The major points of this study are summarized in the Conclusions. The Appendix provides a brief survey of the GR formalism our derivations are based upon.

\section{Maxwell equations and particle and spin dynamics}

\subsection{ Maxwell  equations}

 The local
coordinates are denoted as $x^{\mu},\,\mu=0,1,2,3=(0,i)$, $i=1,2,3$. 
Let the electromagnetic field (2-form) be expressed through 4-potential $A_{\mu}$ (1-form) in local coordinates as:
\begin{gather}
F_{\mu\nu}=\partial_{\mu}A_{\nu}-\partial_{\nu}A_{\mu}.
\label{M10}
\end{gather}
Then the homogeneous Maxwell equations
\begin{gather}
\varepsilon_{\mu\nu\lambda\rho}\partial_{\nu}F_{\lambda\rho}=0
\label{M20}
\end{gather} 
are satisfied automatically. The field (\ref{M10}) is a holonomic one. 

We have also  the inhomogeneous Maxwell equations in the local coordinates:
\begin{gather}
\frac{1}{\sqrt{-g}}\partial_{\nu}\left(\sqrt{-g}F^{\nu\mu}\right)=4\pi J^{\mu},  
\nonumber \\ 
g\equiv\det g_{\mu\nu},  \quad F^{\mu\nu}=g^{\mu\lambda}g^{\nu\rho}F_{\lambda\rho}.
\label{M30}
\end{gather}

The electric and magnetic fields in ONB are defined by the usual rules (see Appendix D):
\begin{gather}
{\bf E}^{\alpha}=F^{\alpha 0}=e^{\alpha}_{\mu}e^0_{\nu}F^{\mu\nu}=e^{\alpha}_ie^0_0F^{i0}+e^{\alpha}_ie^0_jF^{ij},
\label{M40}
\end{gather}
\begin{gather}
\varepsilon_{\alpha\beta\gamma}{\bf H}^{\gamma}=-F^{\alpha\beta}=-e^{\alpha}_{\mu}e^{\beta}_{\nu}F^{\mu\nu},
\quad \varepsilon_{123}=1.
\label{M50}
\end{gather}

\subsection{The particle and spin dynamics}

Let $\d s\equiv \sqrt{g_{\mu\nu}\d x^{\mu}\d x^{\nu}}$ be infinitely small interval  when particle is moving
along its  world line. We use the standard  notations $D X^{\mu}/\d s\equiv u^{\nu}\nabla_{\nu}X^{\mu}$,
where $u^{\nu}\equiv \d x^{\nu}/\d s$, so that $g_{\mu\nu}u^{\mu}u^{\nu}=1$. According to the definition and
Eqs. (\ref{G65}),  (\ref{G70}) we have
\begin{gather}
\frac{D X^a}{\d s}=\frac{\d X^a}{\d s}+\left(\gamma^a_{b\mu}\frac{\d x^{\mu}}{\d s}\right)X^b=\frac{\d X^a}{\d s}+
\left(\gamma^a_{bc}u^c\right)X^b.
\label{D5}
\end{gather}
Then the equation of motion of a charged particle in the local coordinates has the form
\begin{gather}
mc\frac{D u^{\mu}}{\d s}=\frac{q}{c}F^{\mu\nu}u_{\nu}.
\label{D10}
\end{gather}

Evidently, inside the "freely falling elevator" in the  Riemann normal coordinates (\ref{G160}),
the dynamic equations are of the same form as in a flat space-time with Cartesian coordinates:
\begin{gather}
\frac{\d \gamma}{\d t}=\frac{q}{mc}{\bf E}{\boldsymbol\beta},
\nonumber \\
\frac{\d(\gamma{\boldsymbol\beta}^{\alpha})}{\d t}=
\frac{q}{mc}
\left({\bf E}+[{\boldsymbol\beta},\,{\bf H}]\right)^{\alpha},
\nonumber \\
u^a=\gamma(1,\,{\boldsymbol\beta}),  \quad {\boldsymbol\beta}\equiv{\bf v}/c, \quad
\d s=\frac{c}{\gamma}\d t.
\label{D20N}
\end{gather}
Here $\d t$ is the proper time interval in the laboratory frame of reference in which the particle moves with 
the velocity ${\bf v}$.  In the general case 
\begin{gather}
u^a=e^a_{\mu}\frac{\d x^{\mu}}{\d s}.
\label{D22N}
\end{gather}

 A complete description
of spin precession of relativistic particle with MDM and EDM in the gravity field is found in  works \cite{Orlov2012}-\cite{Vergeles2019}.
Let ${\bf S}$ be the polarization vector in the particle rest frame. In the Riemann coordinates, the precession angular velocity ${\bf \Omega}$ equals
\cite{Frenkel1926}-\cite{Fukuyama2013}
\begin{gather}
\frac{\d{\bf S}}{\d t}=\big[{\boldsymbol\Omega},\,{\bf S}\big], 
\nonumber \\ 
{\boldsymbol\Omega}=\frac{q}{mc}\Bigg\{-\left(G+\frac{1}{\gamma}\right){\bf H}+
\frac{\gamma G({\bf H}{\boldsymbol\beta})}{\gamma+1}{\boldsymbol\beta}+
\left(G+\frac{1}{\gamma+1}\right)[{\boldsymbol\beta},\,{\bf E}]+
\nonumber \\
+\eta_{EDM}\left(-{\bf E}+\frac{\gamma}{\gamma+1}({\bf E}{\boldsymbol\beta}){\boldsymbol\beta}+
[{\bf H},\,{\boldsymbol\beta}] \right)\Bigg\}.
\label{D30N}
\end{gather}
Here
\begin{gather}
\mu=(G+1)\frac{q\hbar}{2mc}, \quad d=\eta_{EDM}\frac{q\hbar}{2mc}
\nonumber
\end{gather}
are MDM and EDM, correspondingly.

A formulation of the spin precession problem demands the orthonormal basis (ONB).
 It is important that the fields ${\bf E}$ and ${\bf H}$ in different ONB differ only by the Lorentz transformation. 
However, when describing dynamics in arbitrary ONB, a connection appears in the equations.
 As a result, Eqs. (\ref{D20N}) take the following form:
\begin{gather}
\frac{\d \gamma}{\d t}=\frac{q}{mc}{\bf E}{\boldsymbol\beta}+c(\gamma_{\alpha 0c}u^c){\boldsymbol\beta}^{\alpha},
\nonumber \\
\frac{\d(\gamma{\boldsymbol\beta}^{\alpha})}{\d t}=
\frac{q}{mc}
\left({\bf E}^{\alpha}+[{\boldsymbol\beta},\,{\bf H}]^{\alpha}\right)+
c\Big((\gamma_{\alpha 0c}u^c)+(\gamma_{\alpha\beta c}u^c){\boldsymbol\beta}^{\beta}\Big).
\label{D20}
\end{gather}

Since the polarisation vector ${\bf S}$ is defined in the particle rest frame, the explicit expression for a contribution of the connection to the precession frequency (\ref{D30N}) for a relativistic particle is too  lengthy to be reproduced here \cite{Vergeles2019}, we  only cite a result for the nonrelativistic case:
\begin{gather}
{\boldsymbol\Omega}^{\alpha}=-\frac{q}{mc}\bigg\{\big(G+1\big){\bf H}+
\eta_{EDM}{\boldsymbol E}\bigg\}^{\alpha}
-\frac{c}{2}\varepsilon_{\alpha\beta\rho}
\gamma_{\beta\rho0}.
\label{D30}
\end{gather}

The general conclusion from this brief discussion is 
that the fields ${\bf E}$ and ${\bf H}$ in ONB defined according to
(\ref{M40}) and (\ref{M50}) possess all the dynamic properties of the electric and magnetic fields, correspondingly.
We use hereafter the ONB (\ref{G50}) which is a physically sensible one.

\section{Magnetic field in the pure electrostatic system in the nonertial frame }

Let’s consider a noninertial space-time with the stationary metric in the framework of Einstein’s general theory of relativity. The stationarity means that all components of the metric
tensor $g_{\mu\nu}(x)$  are time-independent, i.e. $(\partial/\partial x^0)g_{\mu\nu}=0$.
A generic feature of the correspontent metric is nonvanishig off-diagonal elements,
\begin{gather}
g_{0i}\neq 0.
\label{intr10}
\end{gather} 
The outlined coordinate frame is denoted as $K$. In the cases of practical interest treated in Section IV the 
reference frame K is the laboratory frame used in the
description of terrestrial experiments. We use also the inertial reference frame of distant stars $K'$.
The  gravitating body and the corresponding frame $K$ rotate w.r.t. to the frame $K'$ with the  angular velocity ${\boldsymbol\omega}$. Our electrostatic system resides at rest on the rotating body and  
the corresponding electric 4-current, as defined in the frame $K$, is the stationary one:
\begin{gather}
(\partial/\partial x^0)J^{\mu}(x)=0, \quad J^0\neq 0, \quad J^i=0.
\label{intr20}
\end{gather}

In this Section we derive the main result of the paper: {\it For the stationary metric} (\ref{intr10})  {\it in the case of stationary electric 4-current} (\ref{intr20})  {\it the Maxwell equations can not be satisfied with the vanishing magnetic field} (see below Eq. (\ref{H60n})).

A proof of this property proceeds as follows. Note that any antisymmetric field in the three-dimensional space $\sqrt{-g}F^{ij}=-\sqrt{-g}F^{ji}$ can be represented as
\begin{gather}
\sqrt{-g}F^{ij}=\varepsilon_{ijk}\partial_k\psi+\left(\partial_i{\cal A}_j-\partial_j{\cal A}_i\right).
\label{H10n}
\end{gather}
The decomposition (\ref{H10n}) would be unique in three-dimensional Euclidean space for fields decreasing at infinity. As well known, in the rotating frames one faces the formal issue of the horizon. However, in all the cases of practical interest, the charge and current distributions are localized way inside the horizon radius. Consequently, the decomposition (\ref{H10n}) is well defined and unique, and $J_i=0$ entails ${\cal A}_i=0$.

According to the rules (\ref{G64}), the field definition (\ref{M50}) and  the fact that $\tilde{e}^i_0=0$, we obtain:
\begin{gather}
F^{ij}=\tilde{e}^i_a\tilde{e}^j_bF^{ab}=-\varepsilon_{\alpha\beta\gamma}
\tilde{e}^i_{\alpha}\tilde{e}^j_{\beta}{\bf H}^{\gamma}=-\frac{e^0_0}{\sqrt{-g}}
\varepsilon_{ijk}e^{\alpha}_k{\bf H}^{\alpha}.
\label{H20n}
\end{gather}
Here we  used one of  the relations (\ref{H25n}).
 Making use of (\ref{H10n}) (with ${\cal A}_i=0$) in (\ref{H20n}) yields
\begin{gather}
{\bf H}^{\alpha}=-(e^0_0)^{-1}\tilde{e}^i_{\alpha}\partial_i\psi.
\label{H30n}
\end{gather}

Now we turn to the  homogeneous Maxwell equations (\ref{M20}).
We express the  holonomic field $F_{\mu \nu}$  in terms of the electric  and  magnetic fields:
\begin{gather}
F_{i0}=e^a_ie^b_0F_{ab}=e^0_0e^{\alpha}_iF_{\alpha0}=
-e^0_0e^{\alpha}_i{\bf E}^{\alpha}=\partial_iA_0, \quad
 \mbox{or}
 \quad {\bf E}^{\alpha}=-\frac{\tilde{e}^i_{\alpha}}{e^0_0}F_{i0},
\label{H40n}
\end{gather} 
\begin{gather}
F_{ij}=e^a_ie^b_jF_{ab}=-\varepsilon_{\alpha\beta\gamma}e^{\alpha}_ie^{\beta}_j{\bf H}^{\gamma}-
\left(e^0_je^{\alpha}_i-e^0_ie^{\alpha}_j\right){\bf E}^{\alpha}.
\label{H50n}
\end{gather} 
Eq. (\ref{M20}) with $\mu=0$ imply the identity $\varepsilon_{ijk}\partial_kF_{ij}=0$. Therefore,
applying the operator $\varepsilon_{ijk}\partial_k$ to the Eq. (\ref{H50n}), using (\ref{H30n})
and the identity $\varepsilon_{ijk}\partial_kF_{i0}=-\varepsilon_{ijk}\partial_k\left(
e^0_0e^{\alpha}_i{\bf E}^{\alpha}\right)=0$ (see (\ref{H40n})), we
 obtain the equation 
\begin{gather} 
\partial_i\bigg(\sqrt{-g}\left(e^0_0\right)^{-2}g^{ij}\partial_j\psi\bigg)=
-\varepsilon_{ijk}e^0_0e^{\alpha}_i{\bf E}^{\alpha}\partial_k\left(\frac{e^0_j}{e^0_0}\right).
\label{H60n}
\end{gather} 
Recall that  $e^0_j=g_{0j}/\sqrt{g_{00}}\neq0$  according to (\ref{G130}) and (\ref{intr10}). Therefore
if ${\bf E}\neq0$, then (\ref{H60n}) entails  $\partial_j\psi\neq0$ and,
according to Eq. (\ref{H30n}),  ${\bf H}\neq0$ as well. We shall refer to the field (\ref{H30n}) as the geometric magnetic field $ {\bf H}_{\boldsymbol\omega}$. As we shall see, the geometric magnetic field has its origin in the rotation of the frame $K$, hence the subscript ${\boldsymbol\omega}$.

Next we look into the equation for the electric field.
Let's  express $F^{i0}$ in terms of physical fields:
\begin{gather} 
F^{i0}=\tilde{e}^i_a\tilde{e}^0_bF^{ab}=\tilde{e}^0_0\tilde{e}^i_{\alpha}{\bf E}^{\alpha}-
\varepsilon_{\alpha\beta\gamma}\tilde{e}^i_{\alpha}\tilde{e}^0_{\beta}{\bf H}^{\gamma}
=\frac{1}{e^0_0}\tilde{e}^i_{\alpha}{\bf E}^{\alpha}-\frac{1}{\sqrt{-g}e^0_0}\varepsilon_{ijk}e^0_j\partial_k\psi.
\label{H70n}
\end{gather} 
The substitution of the right-hand side of (\ref{H70n}) into Eq. (\ref{M30}) with $\mu=0$ leads to 
\begin{gather}
\partial_i\left(\frac{\sqrt{-g}}{e^0_0}\tilde{e}^i_{\alpha}{\bf E}^{\alpha}\right)-
\varepsilon_{ijk}\partial_i\left(\frac{e^0_j}{e^0_0}\right)\cdot\partial_k\psi=
4\pi\sqrt{-g}J^0.
\label{H80n}
\end{gather}
Here we have used representation (\ref{H30n}) and one of the relations (\ref{H25n}).

The system of equations  (\ref{H30n}), (\ref{H60n}) and (\ref{H80n})  is complete  and exact. The electric field ${\bf E}$  is expressed in terms of the potential $A_0$ with the help of relation (\ref{H40n}).

To be specific, we describe the curved space-time of the rotating body by the Kerr  metric described in Appendix B.  We report here the iterative solutions for the electromagnetic potentials $A_{0}$ and $\psi$ and the corresponding fields, treating the angular velocity of the rotating body  ${\boldsymbol{\omega}}$ and its gravitational radius $r_g$ as small parameters. Making use of (\ref{G120})-(\ref{G150}), we rewrite Eqs.  (\ref{H30n}), (\ref{H60n}) and (\ref{H80n}) keeping the terms of the order of   ${\boldsymbol{\omega}}$,
$r_g$, $r_g{\boldsymbol{\omega}}$, ${\boldsymbol{\omega}}\otimes{\boldsymbol{\omega}}$. 

 There emerges a simple hierarchy of corrections in powers of ${\boldsymbol{\omega}}$. We are interested in the electrostatic system with the initial potential $A_0^{(0)}$ and the corresponding Minkowski space defined electric field  ${\boldsymbol{E}}^{(0)}=-\nabla A_0^{(0)}$. Concerning the geometric magnetic field, the crucial points are Eq. (B3) for the off diagonal $g_{0i}$, Eq. (B5) for $e^0_i$ and Eq. (B6) for $\tilde{e}^0_{\alpha}$. The three related  quantities are all proportional to the angular velocity of the rotating body.  Consequently  $e^0_j/e^0_0=\O({\boldsymbol\omega})$ and, according to (\ref{H60n}), the expansion for the magnetic potential $\psi$ starts with the linear term $\psi^{(1)}=\O({\boldsymbol{\omega}})$. 
 	
 Now we proceed to the electric field. According to the expansion of the Kerr metric in Appendix B, we have
\begin{gather}
\frac{\sqrt{-g}\tilde{e}^i_{\alpha}}{e^0_0}=\left(1+\frac{r_g}{R}\right)\delta^i_{\alpha}
+\left(\frac{[{\boldsymbol\omega},\,{\bf R}]^2}{2c^2}\delta^i_{\alpha}-
\frac{[{\boldsymbol\omega},\,{\bf R}]^i[{\boldsymbol\omega},\,{\bf R}]^{\alpha}}{2c^2}\right)\, ,
\label{H90n}
\end{gather}
which does not contain the linear term.  Then, equations (\ref{H80n}) and (\ref{H90n}) tell that the electric field  acquires the first correction only to the second order in ${\boldsymbol{\omega}}$, i.e., $ {\bf E}^{(1)}=0, \,{\bf E}^{(2)}\neq 0$. In the due turn,  Eq. (\ref{H60n})  guarantees that the quadratic correction to the magnetic potential vanishes: $\psi^{(2)}=0$. 

Going back to the magnetic potential $\psi$, we make use of 
\begin{gather}
\frac{\tilde{e}^i_{\alpha}}{e^0_0}=\delta^i_{\alpha}+
\left(\frac{[{\boldsymbol\omega},\,{\bf R}]^2}{2c^2}\delta^i_{\alpha}-
\frac{[{\boldsymbol\omega},\,{\bf R}]^i[{\boldsymbol\omega},\,{\bf R}]^{\alpha}}{2c^2}\right).
\label{H100n}
\end{gather} 
and invoke the relations 
\begin{gather}
\frac{\sqrt{-g}}{\left(e^0_0\right)^2}g^{ij}=-\left(1+\frac{r_g}{R}\right)\delta^{ij}, \quad
e^0_0e^{\alpha}_i=\delta^{\alpha}_i+\O({\boldsymbol{\omega}}\otimes{\boldsymbol{\omega}}), 
\nonumber \\
\frac{e^0_j}{e^0_0}=-\Bigg\{1+\frac{r_g}{R}\left(2-\frac{I R^2_{\oplus}}{R^2}\right)\Bigg\}
\frac{[{\boldsymbol\omega},\,{\bf R}]^j}{c},
\label{H120n}
\end{gather} 
Then Eq. (\ref{H60n}) takes the final form 
\begin{gather}
\partial_i\bigg\{\left(1+\frac{r_g}{R}\right)\partial_i\psi\bigg\}
\nonumber \\
=\varepsilon_{ijk}{\bf E}^{(0)i}
\partial_j\Bigg\{\left[1+\frac{r_g}{R}\left(2-\frac{I R^2_{\oplus}}{R^2}\right)\right]\frac{[{\boldsymbol\omega},\,{\bf R}]^k}{c}\Bigg\}.
\label{H130n}
\end{gather} 

A departure of the metric of the curved space-time from the Minkowski one changes the relationship between the potentail $A_{0}$ and the charge distribution and the relationship between the electric field ${\bf E}^{(2)}$ and the gradient of $A_0^{(2)}$. Although these second order corrections are unlikely to be of any practical significance in the terrestrial laboratories, we cite them for the sake of completeness. To the zeroth order in ${\bf \omega}$, but keeping the terms linear in $r_g$, we have
\begin{gather}
-\nabla\bigg\{\left(1+\frac{r_g}{R}\right)\nabla A^{(0)}_0\bigg\}=4\pi\left(1+\frac{r_g}{R}\right)J^0,.
\label{H160n}
\end{gather}
After some algebra, Eq. (\ref{H80n}) with the account for  Eqs. (\ref{H90n}), (\ref{H120n}) yields the equation for the second order correction to the scalar potential,  
\begin{gather}
-\Delta A^{(2)}_0=\frac{1}{c^2}\nabla\bigg\{[{\boldsymbol\omega},{\bf R}]^2\nabla A^{(0)}_0-
\left([{\boldsymbol\omega},{\bf R}]\nabla A_0^{(0)}\right)[{\boldsymbol\omega},{\bf R}]\bigg\}
-\frac{2}{c}({\boldsymbol\omega}\nabla\psi).
\label{H170n}
\end{gather}
In this derivation, to the desired accuracy, we  made use of $e^0_j/e^0_0=-[{\boldsymbol\omega},\,{\bf R}]/c$. 
Note how the second order correction to the electric potential couplles to the first order potential for the geometric magnetic field. 

Finally, upon using  Eq. (\ref{H100n}) and gathering together all second order corrections, the Eq. (\ref{H40n}) yields the second order correction to the electric field 
\begin{gather}
{\bf E}^{(2)}=-\nabla A_0^{(2)}-\frac{[{\boldsymbol\omega},\,{\bf R}]^2}{2c^2}\nabla A_0^{(0)}
+\left([{\boldsymbol\omega},\,{\bf R}]\nabla A_0^{(0)}\right)\frac{[{\boldsymbol\omega},\,{\bf R}]}{2c^2}.
\label{H150n}
\end{gather} 
Apart from the gradient of the second order scalar potential, here emerge corrections quadratic in the angular velocity ${\bf \omega}$. The system of equations  (\ref{H130n})-(\ref{H150n}) is complete  to the desired accuracy. 
 
The rotating body of the practical interest is the Earth. It is the case of weak gravity. On the terrestrial surface  
$r_g/R_{\oplus}\sim 10^{-9}$  and $\omega R_{\oplus} \sim 1.5\cdot 10^{-6}$. Hence we keep the terms $\O(|{\boldsymbol\omega}|)$ and neglect the former corrections. In this approximation
equations  (\ref{G120})-(\ref{G150}) simplify to  
\begin{gather}
g_{00}=g^{00}=1, \quad g_{0i}=g^{0i}=-\frac{\big[{\boldsymbol\omega},\,{\bf R}]^i}{c},
\nonumber \\
g_{ij}=g^{ij}=-\delta^{ij}, \quad g=-1,
\nonumber \\
e^0_0=1, \quad e^0_i=-\frac{\big[{\boldsymbol\omega},\,{\bf R}\big]^i}{c},
\quad e^{\alpha}_i=\delta^{\alpha}_i, \quad e^{\alpha}_0=0,
\nonumber \\
\tilde{e}^0_0=1, \quad \tilde{e}^0_{\alpha}=\frac{\big[{\boldsymbol\omega},\,{\bf R}\big]^{\alpha}}{c},
\quad \tilde{e}_{\alpha}^i=\delta_{\alpha}^i, \quad \tilde{e}^i_0=0,
\nonumber \\
\frac{c}{2}\varepsilon_{\alpha\beta\rho}\gamma_{\beta\rho 0}={\boldsymbol\omega}^{\alpha}.
\label{H90}
\end{gather}
The second order corrections to the electric potential and electric field can be neglected and we have the familiar ${\bf E}=-\nabla A_0$ and  the Poisson equation $\div{\bf E}=-\Delta A_0=4\pi J^0$, while the Poisson equation for the potential of the geometric magnetic field takes a simple form, 
\begin{gather}
\Delta\psi=\frac{2}{c}({\boldsymbol\omega}{\bf E})\,,
\label{H140}
\end{gather}
to be used in the subsequent analysis of terrestrial experiments.

\section{Manifestations of the geometric magnetic field}

\subsection{False EDM signal in the neutron EDM experiments }

Here we comment on the possible implications of the geometrical magnetic field for the neutron EDM experiments. The fundamental observable is the change of the Larmor precession frequency 
\begin{equation}
f_n =\frac{1}{\pi\hbar}|\mu_n{\bf B} +d_n{\bf E}| \label{eq:Larmor}
\end{equation}
from the parallel to antiparallel uniform fields. The EDM is extracted from the frequency shift
\begin{equation} 
d_n = \frac{\pi \hbar \Delta f}{2|{\bf E}|} . 
\label{eq:EDM extraction}
\end{equation}
The implicit assumption is that flipping the electric field does not change the magnetic one. Our point is that this is not the case with the geometric magnetic field.

In practice the electric field is generated in the plane capacitor with the gap much smaller than the plate size. The relevant solution of the one-dimensional problem for the geometric magnetic field proceeds as follows. In the gap in between the plates we have 
\begin{gather}
{\bf E}_0=(0,\,0,\,{\cal E}_0)=-\nabla A_0, \quad A_0=-{\cal E}_0z.
\label{IIiE10}
\end{gather}
Now we solve the Poisson equation (\ref{H140}) for the magnetic potential, representing the electric field through $A_0(z)$ and integrating by parts:
\begin{gather}
\psi(z)=\frac{2{\boldsymbol\omega}_z}{c}\int\d z'\Delta^{-1}(z-z'){\cal E}_0
=-\frac{2{\boldsymbol\omega}_z}{c}\int\d z'\frac{\d}{\d z}\Delta^{-1}(z-z')A_0(z'),
\label{IIiE20}
\end{gather}
where $\Delta^{-1}(z)=|z|/2$ is the inverse to the Laplace operator. One more differentiation yields \begin{gather}
{\bf H}_{\boldsymbol\omega}=-\nabla\psi(z)
=\left(0,\,0,\,\frac{2{\boldsymbol\omega}_z}{c}A_0(z)\right)=
\left(0,\,0,\,-\frac{2{\boldsymbol\omega}_z{\cal E}_0}{c}z\right)=
-\frac{2{\boldsymbol\omega}_zz}{c}{\bf E}_0.
\label{IIiE30}
\end{gather}
The geometric  field ${\bf H}_{\boldsymbol\omega}$ is parallel to the external electric field ${\bf E}_0$. Its salient feature is the nonvanishing constant gradient
\begin{equation}
\frac{\d{\bf H}_{\boldsymbol\omega}}{\d z} = -\frac{2{\boldsymbol\omega}_z}{c}{\bf E}_0 
\label{eq:Gradient}
\end{equation}
At the mid plane the geometric field vanishes: ${\bf H}_{\boldsymbol\omega}(0)=0$.

The crucial feature of the neutron EDM experiments is the comagnetometry: one measures the neutron spin precession frequency with respect to that of the mercury comagnetometer. The mercury atoms are evenly distributed in the volume of the neutron storage cell and the average geometric magnetic field acting on the mercury comagnetometer vanishes: $\langle {\bf H}_{\boldsymbol\omega}^{(Hg)}\rangle =
{\bf H}_{\boldsymbol\omega}(0) =0$.
The centre of mass of neutrons differs from that of the mercury by the  offset  $\langle z \rangle$, what entails the nonvanishing average geometric magnetic field acting on the magnetic moment of neutrons
\begin{equation}
{\bf H}_{\boldsymbol\omega}^{(n)} = -\frac{2\langle z\rangle{\boldsymbol\omega}_z}{c}{\bf E}_0 .
\label{eq:HgeomNeutron}
\end{equation}
The most important point is that this geometric field changes the sign when the electric field is flipped. The net effect is that the apparent EDM of neutrons, $d_n^{obs}$, as given by the procedure (\ref{eq:EDM extraction}), will acquire the false component, $d_{n}^{obs} = d_n + d_{false}$, where 
\begin{equation}
d_{false} = -\frac{2\langle z \rangle{\boldsymbol\omega}_z}  {c}\mu_n\, . \label{eq:dFalse}
\end{equation}
In the experiment \cite{Pendlebury2015} the neutron center of mass offset was $\langle z \rangle \simeq 2.8$mm, the more recent experiment reports $\langle z \rangle \simeq 3.9$mm. Taking the former, we find $d_{false} \approx 2.5\times  10^{-28}$ e$\cdot$cm. It is still way below the recently reported best upper bound on the neutron EDM, $d_n = (0.0\pm 1.1_{stat} \pm 0.2_{sys})\times 10^{-26}$e$\cdot$cm   
 \cite{Afach2020}, but can become sizeable in the next generation of the neutron EDM experiments aiming at $d_n < 10^{27}$ e$\cdot$cm  \cite{Chupp2019}. With the neutron storage cell of height 12 cm, the geometric magnetic field induced spread of the false EDM within the ensemble of stored neutrons can be as large as 
 \begin{equation}
 \Delta d_{false} = \pm \frac{h{\bf \omega}_z}{c}\mu_n \simeq \pm 5\times 10^{-27} {\text e}\cdot\text{cm} \, . \label{eq:dFalseSpread}
 \end{equation}

\subsection{Geometric  magnetic field as a background in all electric proton  EDM storage rings}

The principal idea of searches for the proton EDM in the all electric ring, run at the so-called magic energy, is to eliminate the magnetic field acting on the proton magnetic moment. Then the sole rotation of the proton spin will be due to interaction of its EDM with the radial electric field that confines protons in the storage ring, and very ambitious sensitivity  to the proton EDM,
\begin{gather}
\eta_{EDM}\sim10^{-15}\, ,
\label{IIE}
\end{gather}
is in sight \cite{Anastopoulos2016,Abusaif2019}.  Here we comment on implications of the geometric magnetic field for such proton EDM experiments. 

The storage ring is a cylinder capacitor with the gap $d$
which is  much smaller compared to the height of cylinders $h$, which in its turn is  much smaller than radii of cylinders $r_{1,2}=\rho\mp d/2$. In view of $d\ll h\ll\rho$ we neglect the dependence on the vertical coordinate and have the two-dimensional geometry. The beam trajectory is in the midplane of the storage ring at the orbit radius $|{\bf r}|=\rho$. The electric field in the gap is given by 
\begin{gather}
{\bf E}_0=-{\cal E}_0\frac{\rho{\bf r}}{r^2} = -\nabla A_{0}(r), \quad\quad A_0(r) = {\cal E}_0 \rho \ln\frac{r}{\rho}\, .
\label{IIE10}
\end{gather}

It is instructive to start with the storage ring located on the North pole. From the viewpoint of distant observer in the reference frame K', the ring rotates with the Earth's rotation angular velocity ${\bf \omega}$. The static charges on the two rotating cylinders produce the opposite currents and generate in the gap the magnetic fields of the same sign and magnitude. The net result is the magnetic field
\begin{gather}
{\bf H}'_{\boldsymbol\omega}({\bf r})=\frac{1}{c}\big[{\bf v}({\bf r}),\,{\bf E}_0({\bf r})\big].
\label{IIE20}
\end{gather}
At first sight, it introduces an asymmetry between the  clockwise and anticlockwise beams in the all electric storage ring, parasitic from the viewpoint of searches for the proton EDM. However, it is basically the motional magnetic field and, to the experimenter in the terrestrial laboratory K, it vanishes entirely. Such an exact cancellation only holds at the North and South poles, and at an  arbitrary  latitude it does not work. In the generic case, the result  (\ref{IIE20}) for the magnetic field  suggests the small parameter $\eta_{\boldsymbol\omega}={|{\boldsymbol\omega}|\rho}/{c}$, similar to that appearing in Eq. (\ref{eq:dFalse}). For the storage ring of radius $\rho \sim 50$m we have 
\begin{gather}
\eta_{\boldsymbol\omega}=\frac{|{\boldsymbol\omega}|\rho}{c}\sim 10^{-11}.
\label{IIE30}
\end{gather}
which is some four orders in magnitude larger than the target value $\eta_{EDM}\sim 10^{-15}$.

Now we solve for the geometric magnetic field following the formalism of Section III. The electric field (\ref{IIE10}) suggests  for the magnetic potential $\psi$ the Ansatz
\begin{gather}
\psi=f(r)\cdot({\boldsymbol\omega_t}{\bf r})\, ,
\label{IIE40}
\end{gather}
where ${\boldsymbol\omega}_t$ is a projection of the Earth's angular velocity onto the ring plane. A generic solution to Eq. (\ref{H140}) is
\begin{gather}
f(r)=-\frac{{\cal E}_0\rho}{c}\left(\ln\frac{r}{\rho}+\zeta\right) = -\frac{A_o(r)}{c} - \frac{{\cal E}_0\rho}{c}\zeta,
\label{IIE70}
\end{gather}
and 
\begin{gather}
{\bf H}_{\boldsymbol\omega}^i= \frac{2{\boldsymbol\omega}_t^j}{c}A_{ij}({\bf r}) =\Bigg\{\frac{A_0}{c}\delta_{ij}
+\frac{{\cal E}_0\rho}{c}\bigg[\left(\zeta-1/2\right)\delta_{ij}
+\frac12(\delta_{ij}-2{\bf n}_i{\bf n}_j)\bigg]\Bigg\}
{\boldsymbol \omega}_t^j,
\label{IIE80}
\end{gather}
where ${\bf n}= {\bf r}/r$ .

The constant $\zeta$ is fixed by the boundary condition that the electric potential $A_{0}$ vanishes rapidly beyond the capacitor, so that in the integral representation for $\psi$ one can perform the integration by parts:
\begin{gather}
\psi({\bf r})=\frac{2{\boldsymbol\omega}_t^j}{c}\int\d^{(2)}y\,\Delta^{-1}({\bf r}-{\bf y}){\bf E}^j_0({\bf y})
=-\frac{2{\boldsymbol\omega}_t^j}{c}\int\d^{(2)}y\,\partial_j\Delta^{-1}({\bf r}-{\bf y})A_0(\bf y), ,
\label{IIE90}
\end{gather}
where $\Delta^{-1}({\bf r})=(1/2\pi)\ln|{\bf r}|$ is the inverse to the Laplace operator. The resulting equation for the symmetric matrix $A_{ij}({\bf r})$
\begin{gather}
A_{ij}({\bf r})=\int\d^{(2)}y\,\partial_i\partial_j\Delta^{-1}({\bf r}-{\bf y})A_0({\bf y}).
\label{IIE100}
\end{gather}
entails 
\begin{gather}
\tr A({\bf r})=A_0({\bf r}).
\label{IIE110}
\end{gather}
Hence the expansion of $A_{ij}$ into irreducible tensor structures is of the form  
\begin{gather}
A_{ij}({\bf r})=\frac{1}{2}\delta_{ij}A_0({\bf r}) + \sigma({\bf r})(\delta_{ij}-
2{\bf n}^i{\bf n}^j) \, ,
\label{IIE120}
\end{gather}
and a comparison to (\ref{IIE80}) gives immediately
\begin{gather}
\sigma({\bf r}) = \frac{1}{4}{\cal E}_0\rho, \quad \zeta = \frac{1}{2}.
\label{IIE130}
\end{gather}
Our final result for the geometric magnetic field in the gap of the storage ring is
\begin{gather}
{\bf H}_{\boldsymbol\omega}^i=\frac{{\cal E}_0\rho}{c}\Bigg\{\ln\left(\frac{r}{\rho}\right)\cdot\delta_{ij}+
\left(\frac12\delta_{ij}-{\bf n}_i{\bf n}_j\right)\Bigg\}
{\boldsymbol \omega}_t^j\simeq  \frac{{\cal E}_0\rho}{2c}
\left(\delta_{ij}-2{\bf n}_i{\bf n}_j\right){\boldsymbol \omega}_t^j,
\label{IIE130}
\end{gather}
where in the last step we neglected $|\log(r/\rho)| < d/(2\rho) \ll 1$.

The background magnetic fields are of prime concern to the planned searches for the proton EDM in the all electric magic storage rings, for a detailed  discussion see the recent monographic document by the CPEDM (Charged Particles EDM) collaboration \cite{Abusaif2019}. Important virtue of the all electric rings is a cancellation of many systematic effects when one compares spin rotations of simultaneously stored clockwise (CW) and anticlockwise (ACW) rotating protons. The magnetic Lorentz forces split the orbits of the CW and ACW beams. As discussed extensively in \cite{Abusaif2019}, the modern techniques allow a very strong, but as yet incomplete,  screening of the Earth's magnetic field.  Despite much work, reported in \cite{Abusaif2019}, the analysis of the magnetic imperfection effects is still in the formative stage.

The Earth's magnetic field ${\bf H}_\oplus$ and the geometric magnetic field ${\bf H}_{\bf \omega}$ do differ markedly. In contrast to the Earth's magnetic field, the  geometric one is not subject to screening by magnetic shields. On the scale of the storage ring, the Earth's magnetic field can be regarded as a uniform one and has the constant projection onto the ring plane. It is pointing along the (magnetic) meridian, which we take for the y-axis : ${\bf H}_\oplus ^t = (0, H_\oplus ^t )$. In contrast to that, the geometric magnetic field has the quadrupole-like behaviour along the particle orbit, ${\bf H}_{\bf \omega} = H_{\bf \omega}(\sin 2\theta, \cos 2\theta) $. The angular position of the particle in a ring,  $\theta$, is defined by ${\bf n} = (\cos \theta, -\sin \theta)$. 

In the all  electric magnetic rings the most dangerous ones are the radial magnetic fields. In the above two cases they are  equal to  $H_\oplus^{(r)} = ({\bf n}\cdot {\bf H}_\oplus) = -H_\oplus \sin \theta $  and $H_{\bf \omega}^{(r)} = ({\bf n}\cdot {\bf H}_{\bf \omega}) = H_{\bf \omega} \sin \theta $\, 
According to \cite{Abusaif2019}, to the first approximation the rotation of the proton spin is proportional to the one-turn integral  $\oint d\theta H_\oplus^{(r)} $. .Obviously, both the Earth's  and  geometric magnetic fields share the property 
\begin{equation}
\oint d\theta H_\oplus^{(r)}  = \oint d\theta H_{\bf \omega}^{(r)} =0\, . \label{eq:EDMlike}
\end{equation}
The argument of Ref.  \cite{Abusaif2019} about vanishing false EDM signal from is then applicable to the geometric magnetic field as well.

The above consideration is somewhat naive and must be complemented by a consistent  treatment of the spin-orbit coupling, though. Namely, this cancellation of the false EDM effect might become incomplete because of the orbit distortions which are very much distinct in the two cases.  To be on the safe side,  one needs a dedicated analysis of the false spin rotations with simultaneous allowance for the orbit distortions. Furthremore, one needs to pay an attention to  a possible cross talk between the impact of the geometric magnetic field and the residual Earth's magnetic field. It is an important complex issue on its own to be addressed to in the future,  it  goes beyond the scope of the present communication.

\subsection{Geometric magnetic field of the conducting charged sphere}

For the sake of completeness, we comment on the charged sphere at rest in the  rotating system $K$. Inside the sphere we have ${\bf E}=0$  and $A_0=\const$ thereof: 
\begin{gather}
{\bf E}({\bf r})=\left\{
\begin{array}{rl}
\dfrac{Q{\bf r}}{r^3},  &  \mbox{for} \quad r>a, \\  [4mm]
0, & \mbox{for} \quad r<a.
\end{array}  \right.
\label{IE10}
\end{gather}
Here $Q$ is the charge of the sphere, and the  radius-vector ${\bf r}=0$ at the centre of the  sphere. 

According to (\ref{H140})
\begin{gather}
\psi({\bf r})=-\frac{\boldsymbol\omega}{2\pi c}\int\d^{(3)}x\frac{1}{|{\bf r}-{\bf x}|}{\bf E}({\bf x})=
-\frac{Q\cdot ({\boldsymbol\omega}{\bf r})}{3c}\left\{
\begin{array}{rl}
\dfrac{3}{r}-\dfrac{a^2}{r^3},  &  \mbox{for} \quad r>a, \\  [4mm]
\dfrac{2}{a}, & \mbox{for} \quad r<a\,,
\end{array}  \right.
\label{IE30}
\end{gather}
and 
\begin{widetext}
\begin{gather}
{\bf H}_{\boldsymbol\omega}({\bf r})=\left\{
\begin{array}{rl}
\dfrac{1}{r^3}\Big\{3({\boldsymbol\mu}{\bf n}){\bf n}-{\boldsymbol\mu}\Big\}+
\dfrac{1}{c}\big[{\bf E}({\bf r}),\,[{\boldsymbol\omega},\,{\bf r}]\big], \quad {\bf E}({\bf r})=\dfrac{Q{\bf r}}{r^3} &  \mbox{for} \quad r>a, \\ [4mm]
\dfrac{2Q{\boldsymbol\omega}}{3ca}, & \mbox{for} \quad r<a.
\end{array}  \right.
\label{IE40}
\end{gather}
\end{widetext}
Here ${\bf n}={\bf r}/{r}$ and
\begin{gather}
{\boldsymbol\mu}=\frac{Qa^2}{3c}{\boldsymbol\omega}.
\nonumber
\end{gather}
is the geometric magnetic moment  of  the conducting charged sphere, induced by the Earth's rotation. The second term in (\ref{IE40}) is the
familiar motional magnetic field in the rotating frame $K$ which is entailed by the electric field in inertial frame $K'$.

\section{Conclusions}

We have shown that in the pure electrostatic systems at rest on the rotating bodies there can exist a geometric magnetic field. From the general relativity point of view, it originates from the nonvanishing off-diagonal elements $g_{0i}$ of the metric tensor which are proportional to the angular velocity of rotation of the gravitating body. We presented several specific examples of the geometric field for different configurations of the static electric field on the rotating body described by the Kerr metric. In the configuration of experimental setups used in the terrestrial searches for the EDM of neutrons, the geometric magnetic field changes the sign when the electric field is flipped. Consequently, its interaction with the magnetic moment of the neutron can imitate the neutron EDM and that can become a sizable background in the next generation of the neutron EDM experiments. We found a fairly large background geometric magnetic field in all electric magic storage rings considered a dedicated machine for searches of the proton EDM. The symmetry properties  of the geometric magnetic field suggest strong cancellations of its contribution to the proton spin rotations. Stil, its impact on the signal of EDM remains an open issue - here one needs a dedicated analysis with full allowance for the spin-orbit dynamics in the storage ring.

\begin{acknowledgments}

We are grateful to A.Ya. Maltsev, A.A. Starobinsky and  S.S. Vergeles for valuable comments and discussions.
This work was carried out as a part of the State Program 0033-2019-0005.

\end{acknowledgments}

\appendix

\section{Geometry}

We consider stationary metric in the reference frame K. Following the textbook \cite{Landau2}, let's diagonalize  this quadratic form:
\begin{gather}
\d s^2=g_{\mu\nu}\d x^{\mu}\d x^{\nu}
=g_{00}\left(\d x^0+\frac{g_{0i}\d x^i}{g_{00}}\right)^2
-\left(-g_{ij}+\frac{g_{0i}g_{0j}}{g_{00}}\right)\d x^i\d x^j 
\nonumber \\
=\eta_{ab}\left(e^a_{\mu}\d x^{\mu}\right)
\left(e^b_{\nu}\d x^{\nu}\right), 
\label{G10}
\end{gather}
where the field $e^a_{\mu}(x)$ is called tetrad,
$a,b,\ldots=0,1,2,3,\, a=(0,\alpha), \,\alpha=1,2,3,\, 
\eta_{ab}=\diag(1,-1,-1,-1))$. Two  infinitesimally close events are simultaneous if 1-form
\begin{gather}
e^0_{\mu}\d x^{\mu}=\sqrt{g_{00}}\left(\d x^0+\frac{g_{0i}\d x^i}{g_{00}}\right)=0.
\label{G20}
\end{gather}
According to (\ref{G10}) and (\ref{G20}) the squared interval between simultaneous events  is
\begin{gather}
-\d s^2=\left(-g_{ij}+\frac{g_{0i}g_{0j}}{g_{00}}\right)\d x^i\d x^j 
=\sum_{\alpha=1}^3\left(e^{\alpha}_i\d x^i\right)\left(e^{\alpha}_j\d x^j\right).
\label{G30}
\end{gather}

The local orthonormal basis (ONB) $\tilde{e}^{\mu}_a(x)$ is defined by the equations
\begin{gather}
e^a_{\mu}(x)\tilde{e}_b^{\mu}(x)=\delta^a_b, \quad g_{\mu\nu}\tilde{e}_a^{\mu}\tilde{e}_b^{\nu}=\eta_{ab}.
\label{G50}
\end{gather}
Since according to (\ref{G30}) 
\begin{gather}
e^{\alpha}_0=0,
\label{G55}
\end{gather}
one readily finds:
\begin{gather}
\tilde{e}^i_0=0, \quad  \tilde{e}^0_0=\left(e^0_0\right)^{-1}, 
\quad
\tilde{e}^i_{\alpha}e_i^{\beta}=\delta^{\beta}_{\alpha},
\quad \tilde{e}^0_{\alpha}=-\left(e^0_0\right)^{-1}e^0_i\tilde{e}^i_{\alpha}.
\label{G60}
\end{gather}

The   rules of the tensor component transitions from coordinate basis to ONB  and vice versa are  standard.
For example
\begin{gather}
X^a=e^a_{\mu}X^{\mu}, \quad X^{\mu}=\tilde{e}^{\mu}_aX^a, \quad
\xi_a=\tilde{e}_a^{\mu}\xi_{\mu}.
\label{G64}
\end{gather}
In ONB the tensor indices are lowered and raised with the help of metric tensors $\eta_{ab}$ and $\eta^{ab}$.
With the above chosen tetrad there is a complete equivaence between  $J^i=0$ and $J^{\alpha}=0$:
\begin{gather}
J^a=e^a_{\mu}J^{\mu}=\left(e^0_0J^0,\,e^{\alpha}_iJ^i\right)=\left(e^0_0J^0,\,0,0,0\right).
\nonumber
\end{gather}

The covariant derivatives $\nabla_{\mu}$ in the coordinate
basis and in ONB are related as
\begin{gather}
\nabla_{\mu}X^a\equiv e^a_{\nu}\nabla_{\mu}X^{\nu}=\partial_{\mu}X^a+\gamma^a_{b\mu}X^b,
\nonumber \\
\nabla_cX^a=\tilde{e}^{\mu}_c\nabla_{\mu}X^a=\tilde{e}^{\mu}_c\partial_{\mu}X^a+\gamma^a_{bc}X^b,
\label{G65}
\end{gather}
where $\gamma^a_{bc}\equiv\tilde{e}^{\mu}_c\gamma^a_{b\mu}$ is the totality of  the connection coefficients,
and 
\begin{gather}
 \gamma_{abc}\equiv\eta_{ad}\gamma^d_{bc}=-\gamma_{bac}.
\label{G70}
\end{gather}
The fact that the connection is free of torsion is fixed by the equations
\begin{gather}
\partial_{\mu}e^a_{\nu}-\partial_{\nu}e^a_{\mu}+\gamma^a_{b\mu}e^b_{\nu}-\gamma^a_{b\nu}e^b_{\mu}=0.
\label{G80}
\end{gather}
The connection coefficients are determined uniquely by Eqs. (\ref{G70}) and (\ref{G80}):
\begin{gather}
\gamma_{abc}=\frac12\left(C_{abc}-C_{bac}-C_{cab}\right),
\nonumber \\
C_{abc}\equiv \eta_{ad}C^d_{bc}, \quad C^a_{bc}=e^a_{\nu,\,i}\left(\tilde{e}^i_b\tilde{e}^{\nu}_c-\tilde{e}^i_c\tilde{e}^{\nu}_b\right).
\label{G90}
\end{gather}

Under the local Lorentz transformation
\begin{gather}
\tilde{e}'^{\mu}_a(x)=\Lambda^b_a(x)\tilde{e}^{\mu}_b(x)
\label{G91}
\end{gather}
the connection 1-form $\gamma^a_{b\mu}\d x^{\mu}$ transforms as follows:
\begin{gather}
\gamma'^a_b=(\Lambda^{-1})^a_c\Lambda^d_b\gamma^c_d+(\Lambda^{-1})^a_c\d \Lambda^c_b.
\label{G92}
\end{gather}
Here  $\Lambda^a_b(x)$ is a local  Lorentz transformation matrix.

\section{Metric and tetrad}

To proceed  further, we need to choose the appropriate metric. Let the rotating reference frame K be  defined for the  Earth rotating with constant angular velocity ${\boldsymbol\omega}$ in the inertial frame  of distant stars $K'$ . The local coordinates,
vectors etc. in $K'$ are denoted as $x^{\prime\mu}$, ${\bf R}'$ and so on. In the frame $K'$ we have the Kerr metric of the rotating  Earth. For the purposes of our analysis it is sufficient yo use a limit of weak gravity  and nonrelativistic rotation velocity and we expand the Kerr metric retaining the terms linear in ${\boldsymbol{\omega}}$ and $r_g$, bilinear in $r_g$ and ${\boldsymbol{\omega}}$
and  quadratic in ${\boldsymbol{\omega}}$:
\begin{gather}
g'_{00}({\bf R}')=1-\frac{r_g}{R'}, 
\nonumber \\ g'_{0i}({\bf R}')=\frac{2kC}{c^3R'^3}\big[{\boldsymbol\omega},\,{\bf R}'\big]^{i}=I\cdot\frac{r_gR_{\oplus}^2}{R^{\prime3}}\frac{\big[{\boldsymbol\omega},\,{\bf R}'\big]^{i}}{c},
\nonumber \\ 
g'_{ij}({\bf R}')=-\left(1+\frac{r_g}{R'}\right)\delta_{ij}, 
\nonumber \\ 
r_g=\frac{2kM_{\oplus}}{c^2},
\label{G100}
\end{gather}
where $R'=|{\bf R}'|$, $k=6,674\cdot10^{-8}\mbox{cm}^3\cdot\mbox{g}^{-1}\cdot\mbox{sec}^{-2}$ is the gravitation constant, $M_{\oplus}$ and  $C=IM_{\oplus}R_{\oplus}^2$ are the Earth mass  and  moment of inertia relative to polar axis, $I=0.3307$, 
$M_{\oplus}=5,972\cdot10^{27}\mbox{g}$, $R_{\oplus}=6,378\cdot10^8\mbox{cm}$, 
 $r_g=2kM_{\oplus}/c^2=0,887\mbox{cm}$.  Next we transform the metric (\ref{G100}) into the metric in the frame $K$ rotating 
with angular velocity ${\boldsymbol\omega}$ relative to the frame $K'$. We take the coordinates in $K$ and $K'$ 
having the same origin at the centre of Earth. 
The local coordinates in the frame $K$ are denoted as $x^{\mu}$ and, by definition, they are connected with coordinates $x^{\prime\mu}$
as follows,
\begin{gather}
\d x^{\prime0}=\d x^0, \quad \d {\bf R}'=\d {\bf R}+[{\boldsymbol{\omega}},\,{\bf R}]\frac{\d x^0}{c},
\quad |{\bf R}'|=|{\bf R}|\, ,  
\label{G110}
\end{gather}
and the metric $g_{\mu\nu}$
in the frame $K$ equals
\begin{gather}
g_{00}=1-\frac{r_g}{R}-\frac{[{\boldsymbol\omega},\,{\bf R}]^2}{c^2}, 
\nonumber \\
 g_{0i}=-\Bigg\{1+\frac{r_g}{R}\left(1-I\cdot\frac{R_{\oplus}^2}{R^2}\right)\Bigg\}\frac{[{\boldsymbol\omega},\,{\bf R}]^i}{c}, 
\nonumber \\
g_{ij}=-\left(1+\frac{r_g}{R}\right)\delta^{ij}.  
\label{G120}
\end{gather}
The inverse metric tensor $g^{\mu\nu}=\eta^{ab}\tilde{e}^{\mu}_a\tilde{e}^{\nu}_b$ is
\begin{gather}
g^{00}=1+\frac{r_g}{R}, 
\nonumber \\
 g^{0i}=-\Bigg\{1+\frac{r_g}{R}\left(1-I\cdot\frac{R_{\oplus}^2}{R^2}\right)\Bigg\}\frac{[{\boldsymbol\omega},\,{\bf R}]^i}{c}, 
\nonumber \\
g^{ij}=-\left(1-\frac{r_g}{R}\right)\delta^{ij}+
\frac{[{\boldsymbol\omega},\,{\bf R}]^i[{\boldsymbol\omega},\,{\bf R}]^j}{c^2}.
\label{G150}
\end{gather}

To the same approximation the tetrad equals   
\begin{gather}
e^0_0=\sqrt{g_{00}}=1-\frac{r_g}{2R}-\frac{[{\boldsymbol\omega},\,{\bf R}]^2}{2c^2}, \quad e^{\alpha}_0=0,
\nonumber \\  
e^0_i=\frac{g_{0i}}{\sqrt{g_{00}}}=-\Bigg\{1+\frac{r_g}{R}\left(\frac32-I\cdot\frac{R_{\oplus}^2}{R^2}\right)\Bigg\}
\frac{[{\boldsymbol\omega},\,{\bf R}]^i}{c},
\nonumber \\
e^{\alpha}_i=\left(1+\frac{r_g}{2R}\right)\delta^{\alpha}_i+
\frac{[{\boldsymbol\omega},\,{\bf R}]^{\alpha}[{\boldsymbol\omega},\,{\bf R}]^i}{2c^2}\, ,
\label{G130}
\end{gather}
and the ONB vector fields are
\begin{gather}
\tilde{e}_0^0=1+\frac{r_g}{2R}+\frac{[{\boldsymbol\omega},\,{\bf R}]^2}{2c^2}, \quad   \tilde{e}_0^i=0,
\nonumber \\
\tilde{e}_{\alpha}^0=\Bigg\{1+\frac{r_g}{R}\left(\frac32-I\cdot\frac{R_{\oplus}^2}{R^2}\right)\Bigg\}
\frac{[{\boldsymbol\omega},\,{\bf R}]^{\alpha}}{c},
\nonumber \\
\tilde{e}^i_{\alpha}=\left(1-\frac{r_g}{2R}\right)\delta_{\alpha}^i-
\frac{[{\boldsymbol\omega},\,{\bf R}]^i[{\boldsymbol\omega},\,{\bf R}]^{\alpha}}{2c^2}.
\label{G140}
\end{gather}

\section{Useful relations}

The following  relations are used  in the   main body of the text:
\begin{gather}
\varepsilon_{abcd}\tilde{e}^{\mu}_a\tilde{e}^{\nu}_b=\frac{1}{\sqrt{-g}}
\varepsilon_{\mu\nu\lambda\rho}e^c_{\lambda}e^d_{\rho}, 
\nonumber \\
\varepsilon_{abcd}\tilde{e}^{\mu}_a\tilde{e}^{\nu}_b\tilde{e}^{\lambda}_c=\frac{1}{\sqrt{-g}}
\varepsilon_{\mu\nu\lambda\rho}e^d_{\rho},
\nonumber
\end{gather}
which imply that
\begin{gather}
\varepsilon_{\alpha\beta\gamma}\tilde{e}^i_{\alpha}\tilde{e}^j_{\beta}=\frac{e^0_0}{\sqrt{-g}}
\varepsilon_{ijk}e^{\gamma}_k, 
\nonumber \\ 
\varepsilon_{\alpha\beta\gamma}\tilde{e}^i_{\alpha}\tilde{e}^0_{\beta}=\frac{1}{\sqrt{-g}}
\varepsilon_{ijk}e^{\gamma}_je^0_k,
\nonumber \\
\varepsilon_{ijk}e^{\alpha}_ie^{\beta}_j=\sqrt{-g}\,\varepsilon_{\alpha\beta\gamma}
\tilde{e}^0_0\tilde{e}^k_{\rho}, 
\nonumber \\
\varepsilon_{\alpha\beta\gamma}\tilde{e}^i_{\alpha}\tilde{e}^0_{\beta}\tilde{e}^j_{\gamma}=\frac{1}{\sqrt{-g}}
\varepsilon_{ijk}e^0_k,
\label{H25n}
\end{gather}
and so forth.

\section{Riemann normal coordinates}

For the correct interpretation of the  electromagnetic fields in a curved space-time it is useful 
to keep in mind the form of  dynamic equations in the Riemann normal coordinates. The 
Riemann normal coordinates $y^{\mu}$ can be introduced in the vicinity of any point $p$, so that
$y^{\mu}(p)=0$ and
\begin{gather}
e^a_{\mu}(y)=\delta^a_{\mu}+\frac16\mR^a_{\nu\lambda\mu}(p)y^{\nu}y^{\lambda}+\O(y^3),
\nonumber \\
g_{\mu\nu}(y)=\eta_{\mu\nu}+\frac13\mR_{\mu\lambda\rho\nu}(p)y^{\lambda}y^{\rho}+\O(y^3), 
\nonumber \\
\eta_{\mu\nu}=\diag(1,\,-1,\,-1,\,-1),
\nonumber \\
\gamma_{ab\,\mu}(y)=\frac12\mR_{ab\,\nu\mu}(p)y^{\nu}+\O(y^2),
\label{G160}
\end{gather}
where $\mR_{ab\,\nu\mu}$ is the Riemann curvature tensor. Since near the Earth surface 
$|\mR_{ab\,\nu\mu}|\sim r_g/R_{\oplus}^3\sim0,5\cdot10^{-26}\mbox{cm}^{-2}$,
one can ignore the space-time curvature in the small vicinity of  point $p$. This  vicinity with
Riemann normal coordinates $\{y^{\mu}\}$ is the mathematical model of a "freely falling elevator". 
 Therefore, when writing differential equations in the center of normal coordinates, the curvature can be neglected. It follows from here that differential dynamic equations in normal Riemann coordinates $\{y^{\mu}\}$
inside the "elevator" have the same form as in the Cartesian coordinates in the Minkowski space.
Obviously, in the  Riemann normal coordinates inside the "elevator", all field components in the ONB $F^{ab}=
e^a_{\mu}e^b_{\nu}F^{\mu\nu}$ coincide with of the same named field components in the Riemann coordinates.
Upon transition to arbitrary ONB, all tensors are transformed in accordance with the usual Lorentz rules, and a connection determined by ONB appears in the equations of motion.


\begin{thebibliography}{11}
\expandafter\ifx\csname natexlab\endcsname\relax\def\natexlab#1{#1}\fi
\expandafter\ifx\csname bibnamefont\endcsname\relax
  \def\bibnamefont#1{#1}\fi
\expandafter\ifx\csname bibfnamefont\endcsname\relax
  \def\bibfnamefont#1{#1}\fi
\expandafter\ifx\csname citenamefont\endcsname\relax
  \def\citenamefont#1{#1}\fi
\expandafter\ifx\csname url\endcsname\relax
  \def\url#1{\texttt{#1}}\fi
\expandafter\ifx\csname urlprefix\endcsname\relax\def\urlprefix{URL }\fi
\providecommand{\bibinfo}[2]{#2}
\providecommand{\eprint}[2][]{\url{#2}}







\bibitem[{\citenamefont{Silenko and Teryaev}(2016)}]{Silenko2007}
\bibinfo{author}{\bibfnamefont{A.J.}~\bibnamefont{Silenko}} \bibnamefont{and}
\bibinfo{author}{\bibfnamefont{O.V.}~\bibnamefont{Teryaev}},
\bibinfo{title}{Equivalence principle and experimental tests of gravitational spin effects},
  \bibinfo{journal}{Physical Review D} \textbf{\bibinfo{volume}{76}},
  \bibinfo{pages}{06110} (\bibinfo{year}{2007}).
  
  
\bibitem[{\citenamefont{Orlov, Flanagan and Semertzidis}(2012)}]{Orlov2012}
\bibinfo{author}{\bibfnamefont{Y.}~\bibnamefont{Orlov}},
\bibinfo{author}{\bibfnamefont{E.}~\bibnamefont{Flanagan}} \bibnamefont{and}
\bibinfo{author}{\bibfnamefont{Y.}~\bibnamefont{Semertzidis}},
\bibinfo{title}{Spin Rotation by Earth's Gravitational Field in a "Frozen-spin" Ring},
\bibinfo{journal}{Phys. Lett.} \textbf{\bibinfo{volume}{A376}}, 
\bibinfo{pages}{2822} (\bibinfo{year}{2012}).


\bibitem[{\citenamefont{Obukhov, Silenko and Teryaev}(2016)}]{Obukhov2016}
\bibinfo{author}{\bibfnamefont{Y.N.}~\bibnamefont{Obukhov}},
\bibinfo{author}{\bibfnamefont{A.J.}~\bibnamefont{Silenko}} \bibnamefont{and}
\bibinfo{author}{\bibfnamefont{O.V.}~\bibnamefont{Teryaev}},
  \bibinfo{journal}{Physical Review D} \textbf{\bibinfo{volume}{94}},
  \bibinfo{pages}{044019} (\bibinfo{year}{2016}).
  


\bibitem[{\citenamefont{Nikolaev, Rathmann, Saleev and  Silenko}(2019)}]{Nikolaev2019}
\bibinfo{author}{\bibfnamefont{N.N.}~\bibnamefont{Nikolaev}},
\bibinfo{author}{\bibfnamefont{F.}~\bibnamefont{Rathmann}},
\bibinfo{author}{\bibfnamefont{A.}~\bibnamefont{Saleev}} \bibnamefont{and}
\bibinfo{author}{\bibfnamefont{A.J.}~\bibnamefont{Silenko}},
\bibinfo{journal}{Invited talk at the 23rd International Spin Physics Symposium - SPIN2018, Ferrara, Italy} \textbf{\bibinfo{volume}{10-14 September, 2018}},
  \bibinfo{pages}{089} (\bibinfo{year}{2019}).
  
  
  
\bibitem[{\citenamefont{Vergeles and Nikolaev}(2019)}]{Vergeles2019}
\bibinfo{author}{\bibfnamefont{S.N.}~\bibnamefont{Vergeles}} \bibnamefont{and}
\bibinfo{author}{\bibfnamefont{N.N.}~\bibnamefont{Nikolaev}},
\bibinfo{title}{Gravitational Effects in Electrostatic Storage Rings and the Search for the Electric Dipole Moments of Charged Particles},
\bibinfo{journal}{JETP} \textbf{\bibinfo{volume}{129}}, \bibinfo{pages}{541-552}
  (\bibinfo{year}{2019}).

\bibitem[{\citenamefont{Abusaif  et al.}(2019)}]{Abusaif2019}
\bibinfo{author}{\bibfnamefont{A.}~\bibnamefont{Abusaif  et al.}},
\bibinfo{title}{Storage Ring to Search for Electric Dipole Moments of Charged Particles - Feasibility Study},
\bibinfo{journal}{CERN-PBC-REPORT-2019-002; e-Print: arXiv} \textbf{\bibinfo{volume}{}}, \bibinfo{pages}{1912.07881 [hep-ex] }(\bibinfo{year}{2019}).

\bibitem[{\citenamefont{ Baker et al.}(2007)}]{Baker2007}
\bibinfo{author}{\bibfnamefont{C.A.}~\bibnamefont{Baker }},
\bibinfo{title}{}
\bibinfo{journal}{Phys. Rev. Lett.} \textbf{\bibinfo{volume}{98}}, \bibinfo{pages}{149102}
  (\bibinfo{year}{2007}).
  








\bibitem[{\citenamefont{Lamoreaux and  Golub}(2007)}]{Lamoreaux2007}
\bibinfo{author}{\bibfnamefont{S.K.}~\bibnamefont{Lamoreaux}} \bibnamefont{and}
\bibinfo{author}{\bibfnamefont{R.}~\bibnamefont{Golub}},
\bibinfo{title}{},
\bibinfo{journal}{Phys. Rev. Lett.} \textbf{\bibinfo{volume}{98}}, \bibinfo{pages}{149101}
  (\bibinfo{year}{2007}).




\bibitem[{\citenamefont{Serebrov, Kolomenskiy, Pirozhkov, Krasnoschekova, Vassiljev, Polyushkin, Lasakov, Murashkina, Solovey, Fomin, Shoka and  Zherebtsov}(2015)}]{Serebrov2015}
\bibinfo{author}{\bibfnamefont{A.P.}~\bibnamefont{Serebrov}},
\bibinfo{author}{\bibfnamefont{E.A.}~\bibnamefont{Kolomenskiy}},
\bibinfo{author}{\bibfnamefont{A.N.}~\bibnamefont{Pirozhkov}},
\bibinfo{author}{\bibfnamefont{I.A.}~\bibnamefont{Krasnoschekova}},
\bibinfo{author}{\bibfnamefont{A.V.}~\bibnamefont{Vassiljev}},
\bibinfo{author}{\bibfnamefont{A.O.}~\bibnamefont{Polyushkin}},
\bibinfo{author}{\bibfnamefont{M.S.}~\bibnamefont{Lasakov}},
\bibinfo{author}{\bibfnamefont{A.N.}~\bibnamefont{Murashkina}},
\bibinfo{author}{\bibfnamefont{V.A.}~\bibnamefont{Solovey}},
\bibinfo{author}{\bibfnamefont{A.K.}~\bibnamefont{Fomin}},
\bibinfo{author}{\bibfnamefont{I.V.}~\bibnamefont{Shoka}} \bibnamefont{and}
\bibinfo{author}{\bibfnamefont{O.M.}~\bibnamefont{Zherebtsov}},
\bibinfo{title}{New searrch for the neutron electric dipol moment with ultracold neutrons at ILL},
\bibinfo{journal}{Physical Review C} 
\textbf{\bibinfo{volume}{92}}, \bibinfo{pages}{055501}
  (\bibinfo{year}{2015}).
    
\bibitem[{\citenamefont{Abel, Ayres, Baker, Ban, Bison, Bodek, Bondar, Crawford, Chiu, Chanel, Chowdhuri, Daum, Dechenaux, Emmenegger, Ferraris-Bouches, Flaux, Geltenbort, Green, Griffith, van der Grinten, Harris, Henneck, Hild, Iaydjiev, Ivanov, Kasprzak, Kermaidic, Kirch, Koch, Komposch, Koss, Kozela, Krempel, Lauss, Lefort, Lemiere, Leredde, Mohanmurthy, Pais, Piegsa, Pignol, Quemener, Rawlik, Rebreyend, Ries, Roccia, Rozpedzic, Schmidt-Wellenburg, Schnabel, Severijns, Virot, Weis,  Wursten, Wyszynski, Zejma and  Zsigmond}(2019)}]{Abel2019}
\bibinfo{author}{\bibfnamefont{C.}~\bibnamefont{Abel}},
\bibinfo{author}{\bibfnamefont{N.J.}~\bibnamefont{Ayres}},
\bibinfo{author}{\bibfnamefont{C.A.}~\bibnamefont{Baker}},
\bibinfo{author}{\bibfnamefont{G.}~\bibnamefont{Ban}},
\bibinfo{author}{\bibfnamefont{G.}~\bibnamefont{Bison}},
\bibinfo{author}{\bibfnamefont{K.}~\bibnamefont{Bodek}},
\bibinfo{author}{\bibfnamefont{V.}~\bibnamefont{Bondar}},
\bibinfo{author}{\bibfnamefont{C.B.}~\bibnamefont{Crawford}},
\bibinfo{author}{\bibfnamefont{P.-J.}~\bibnamefont{Chiu}},
\bibinfo{author}{\bibfnamefont{E.}~\bibnamefont{Chanel}},
\bibinfo{author}{\bibfnamefont{Z.}~\bibnamefont{Chowdhuri}},
\bibinfo{author}{\bibfnamefont{M.}~\bibnamefont{Daum}},
\bibinfo{author}{\bibfnamefont{B.}~\bibnamefont{Dechenaux}},
\bibinfo{author}{\bibfnamefont{S.}~\bibnamefont{Emmenegger}},
\bibinfo{author}{\bibfnamefont{L.}~\bibnamefont{Ferraris-Bouches}},
\bibinfo{author}{\bibfnamefont{P.}~\bibnamefont{Flaux}},
\bibinfo{author}{\bibfnamefont{P.}~\bibnamefont{Geltenbort}},
\bibinfo{author}{\bibfnamefont{K.}~\bibnamefont{Green}},
\bibinfo{author}{\bibfnamefont{W.C.}~\bibnamefont{Griffith}},
\bibinfo{author}{\bibfnamefont{M.}~\bibnamefont{van der Grinten}},
\bibinfo{author}{\bibfnamefont{P.G.}~\bibnamefont{Harris}},
\bibinfo{author}{\bibfnamefont{R.}~\bibnamefont{Henneck}},
\bibinfo{author}{\bibfnamefont{N.}~\bibnamefont{Hild}},
\bibinfo{author}{\bibfnamefont{P.}~\bibnamefont{Iaydjiev}},
\bibinfo{author}{\bibfnamefont{S.N.}~\bibnamefont{Ivanov}},
\bibinfo{author}{\bibfnamefont{M.}~\bibnamefont{Kasprzak}},
\bibinfo{author}{\bibfnamefont{Y.}~\bibnamefont{Kermaidic}},
\bibinfo{author}{\bibfnamefont{K.}~\bibnamefont{Kirch}},
\bibinfo{author}{\bibfnamefont{H.-C.}~\bibnamefont{Koch}},
\bibinfo{author}{\bibfnamefont{S.}~\bibnamefont{Komposch}},
\bibinfo{author}{\bibfnamefont{P.A.}~\bibnamefont{Koss}},
\bibinfo{author}{\bibfnamefont{A.}~\bibnamefont{Kozela}},
\bibinfo{author}{\bibfnamefont{J.}~\bibnamefont{Krempel}},
\bibinfo{author}{\bibfnamefont{B.}~\bibnamefont{Lauss}},
\bibinfo{author}{\bibfnamefont{T.}~\bibnamefont{Lefort}},
\bibinfo{author}{\bibfnamefont{Y.}~\bibnamefont{ Lemiere}},
\bibinfo{author}{\bibfnamefont{A.}~\bibnamefont{Leredde}},
\bibinfo{author}{\bibfnamefont{P.}~\bibnamefont{Mohanmurthy}},
\bibinfo{author}{\bibfnamefont{D.}~\bibnamefont{Pais}},
\bibinfo{author}{\bibfnamefont{F.M.}~\bibnamefont{Piegsa}},
\bibinfo{author}{\bibfnamefont{G.}~\bibnamefont{Pignol}},
\bibinfo{author}{\bibfnamefont{G.}~\bibnamefont{Quemener}},
\bibinfo{author}{\bibfnamefont{M.}~\bibnamefont{Rawlik}},
\bibinfo{author}{\bibfnamefont{D.}~\bibnamefont{Rebreyend}},
\bibinfo{author}{\bibfnamefont{D.}~\bibnamefont{Ries}},
\bibinfo{author}{\bibfnamefont{S.}~\bibnamefont{Roccia}},
\bibinfo{author}{\bibfnamefont{D.}~\bibnamefont{Rozpedzic}},
\bibinfo{author}{\bibfnamefont{P.}~\bibnamefont{Schmidt-Wellenburg}},
\bibinfo{author}{\bibfnamefont{A.}~\bibnamefont{Schnabel}},
\bibinfo{author}{\bibfnamefont{N.}~\bibnamefont{Severijns}},
\bibinfo{author}{\bibfnamefont{R.}~\bibnamefont{Virot}},
\bibinfo{author}{\bibfnamefont{A.}~\bibnamefont{Weis}},
\bibinfo{author}{\bibfnamefont{E.}~\bibnamefont{Wursten}},
\bibinfo{author}{\bibfnamefont{G.}~\bibnamefont{ Wyszynski}},
\bibinfo{author}{\bibfnamefont{J.}~\bibnamefont{ Zejma}} \bibnamefont{and}
\bibinfo{author}{\bibfnamefont{G.}~\bibnamefont{ Zsigmond}},
\bibinfo{title}{Magnetic-field uniformity in neutron electric-dipole-moment experiments},
\bibinfo{journal}{Physical Review A} 
\textbf{\bibinfo{volume}{99}}, \bibinfo{pages}{042112}
  (\bibinfo{year}{2019}).
  
\bibitem[{\citenamefont{Chupp, Fierlinger, Ramsey-Musolf, Singh}(2019)}]{Chupp2019}
\bibinfo{author}{\bibfnamefont{Timothy }~\bibnamefont{Chupp}},
\bibinfo{author}{\bibfnamefont{Peter}~\bibnamefont{Fierlinger}},
\bibinfo{author}{\bibfnamefont{ Michael }~\bibnamefont{Ramsey-Musolf}} \bibnamefont{and}
\bibinfo{author}{\bibfnamefont{Jaideep}~\bibnamefont{Singh }},
\bibinfo{title}{Electric dipole moments of atoms, molecules, nuclei, and particles},
\bibinfo{journal}{Rev.Mod.Phys. } 
\textbf{\bibinfo{volume}{91}}, \bibinfo{pages}{015001}
(\bibinfo{year}{2019}).






\bibitem[{\citenamefont{Anastassopoulos  et al.}(2016)}]{Anastopoulos2016}
\bibinfo{author}{\bibfnamefont{V.}~\bibnamefont{Anastassopoulos  et al.}},
\bibinfo{journal}{Rev. Sci. Instrum.} \textbf{\bibinfo{volume}{87}}, \bibinfo{pages}{115116}
(\bibinfo{year}{2016}).
 
 





 


  
  
  

 

  











\bibitem[{\citenamefont{Frenkel}(1926)}]{Frenkel1926}
\bibinfo{author}{\bibfnamefont{J.}~\bibnamefont{Frenkel}},
  \bibinfo{journal}{Z. Phys.} \textbf{\bibinfo{volume}{37}},
\bibinfo{pages}{243}  (\bibinfo{year}{1926}).


\bibitem[{\citenamefont{Bargmann, Michel and Telegdi}(1959)}]{Bargmann1959}
\bibinfo{author}{\bibfnamefont{V.}~\bibnamefont{Bargmann}},
\bibinfo{author}{\bibfnamefont{L.}~\bibnamefont{Michel}} \bibnamefont{and}
\bibinfo{author}{\bibfnamefont{V.L.}~\bibnamefont{Telegdi}},
  \bibinfo{journal}{Phys. Rev. Lett.} \textbf{\bibinfo{volume}{2}},
  \bibinfo{pages}{435} (\bibinfo{year}{1959}).

\bibitem[{\citenamefont{Khriplovich and Pomeransky}(1998)}]{Khriplovich1998}
\bibinfo{author}{\bibfnamefont{I.B.}~\bibnamefont{Khriplovich}} \bibnamefont{and}
\bibinfo{author}{\bibfnamefont{A.A.}~\bibnamefont{Pomeransky}},
\bibinfo{title}{Equations of Motion for Spinning Relativistic Particle  in External Fields},
  \bibinfo{journal}{J. Exp. Teor. Fis.} \textbf{\bibinfo{volume}{113}},
  \bibinfo{pages}{1537} (\bibinfo{year}{1998}).

\bibitem[{\citenamefont{Pomeransky, Senkov and Khriplovich}(2000)}]{Pomeransky2000}
\bibinfo{author}{\bibfnamefont{A.A.}~\bibnamefont{Pomeransky}},
\bibinfo{author}{\bibfnamefont{R.A.}~\bibnamefont{Senkov }} \bibnamefont{and}
\bibinfo{author}{\bibfnamefont{I.B.}~\bibnamefont{Khriplovich}},
\bibinfo{title}{Spinning Relativistic Particles  in External Fields},
\bibinfo{journal}{Phys. Usp.} \textbf{\bibinfo{volume}{43}}, 
\bibinfo{pages}{1055} (\bibinfo{year}{2000}).




\bibitem[{\citenamefont{Fukuyama and Silenko}(2013)}]{Fukuyama2013}
\bibinfo{author}{\bibfnamefont{T.}~\bibnamefont{Fukuyama}} \bibnamefont{and}
\bibinfo{author}{\bibfnamefont{A.J.}~\bibnamefont{Silenko}},
\bibinfo{title}{Derivation of Generalized Thomas-Bargmann-Michel-Telegdi Equation for a Particle with Electric Dipol Moment},
\bibinfo{journal}{Int. J. Mod. Phys. } \textbf{\bibinfo{volume}{A28}},
\bibinfo{pages}{1350147}  (\bibinfo{year}{2013}).


\bibitem[{\citenamefont{Pendlebury, Afach, Ayres, Baker, Ban, Bison, Bodek, Burghoff, Geltenbort, Green, Griffith, van der Grinten, Grujic, Harris, Helaine, Iaydjiev, Ivanov, Kasprzak, Kermaidic, Kirch, Koch, Komposch, Kozela, Krempel, Lauss, Lefort, Lemiere, May, Musgrave, Naviliat-Cuncic, Piegsa, Pignol, Prashanth, Quemener, Rawlik, Rebreyend, Richardson, Ries, Roccia, Rozpedzic, Schnabel, Schmidt-Wellenburg, Severijns, Shiers, Thome, Weis, Winston, Wursten, Zejma and  Zsigmond}(2015)}]{Pendlebury2015}
\bibinfo{author}{\bibfnamefont{J.M.}~\bibnamefont{Pendlebury}},
\bibinfo{author}{\bibfnamefont{S.}~\bibnamefont{Afach}},
\bibinfo{author}{\bibfnamefont{N.J.}~\bibnamefont{Ayres}},
\bibinfo{author}{\bibfnamefont{C.A.}~\bibnamefont{Baker}},
\bibinfo{author}{\bibfnamefont{G.}~\bibnamefont{Ban}},
\bibinfo{author}{\bibfnamefont{G.}~\bibnamefont{Bison}},
\bibinfo{author}{\bibfnamefont{K.}~\bibnamefont{Bodek}},
\bibinfo{author}{\bibfnamefont{M.}~\bibnamefont{Burghoff}},
\bibinfo{author}{\bibfnamefont{P.}~\bibnamefont{Geltenbort}},
\bibinfo{author}{\bibfnamefont{K.}~\bibnamefont{Green}},
\bibinfo{author}{\bibfnamefont{W.C.}~\bibnamefont{Griffith}},
\bibinfo{author}{\bibfnamefont{M.}~\bibnamefont{van der Grinten}},
\bibinfo{author}{\bibfnamefont{Z.D.}~\bibnamefont{Grujic}},
\bibinfo{author}{\bibfnamefont{P.G.}~\bibnamefont{Harris}},
\bibinfo{author}{\bibfnamefont{V.}~\bibnamefont{Helaine}},
\bibinfo{author}{\bibfnamefont{P.}~\bibnamefont{Iaydjiev}},
\bibinfo{author}{\bibfnamefont{S.N.}~\bibnamefont{Ivanov}},
\bibinfo{author}{\bibfnamefont{M.}~\bibnamefont{Kasprzak}},
\bibinfo{author}{\bibfnamefont{Y.}~\bibnamefont{Kermaidic}},
\bibinfo{author}{\bibfnamefont{K.}~\bibnamefont{Kirch}},
\bibinfo{author}{\bibfnamefont{H.-C.}~\bibnamefont{Koch}},
\bibinfo{author}{\bibfnamefont{S.}~\bibnamefont{Komposch}},
\bibinfo{author}{\bibfnamefont{A.}~\bibnamefont{Kozela}},
\bibinfo{author}{\bibfnamefont{J.}~\bibnamefont{Krempel}},
\bibinfo{author}{\bibfnamefont{B.}~\bibnamefont{Lauss}},
\bibinfo{author}{\bibfnamefont{T.}~\bibnamefont{Lefort}},
\bibinfo{author}{\bibfnamefont{Y.}~\bibnamefont{ Lemiere}},
\bibinfo{author}{\bibfnamefont{D.J.R.}~\bibnamefont{ May}},
\bibinfo{author}{\bibfnamefont{M.}~\bibnamefont{Musgrave}},
\bibinfo{author}{\bibfnamefont{O.}~\bibnamefont{Naviliat-Cuncic}},
\bibinfo{author}{\bibfnamefont{F.M.}~\bibnamefont{Piegsa}},
\bibinfo{author}{\bibfnamefont{G.}~\bibnamefont{Pignol}},
\bibinfo{author}{\bibfnamefont{P.N.}~\bibnamefont{ Prashanth}},
\bibinfo{author}{\bibfnamefont{G.}~\bibnamefont{Quemener}},
\bibinfo{author}{\bibfnamefont{M.}~\bibnamefont{Rawlik}},
\bibinfo{author}{\bibfnamefont{D.}~\bibnamefont{Rebreyend}},
\bibinfo{author}{\bibfnamefont{J.D.}~\bibnamefont{Richardson}},
\bibinfo{author}{\bibfnamefont{D.}~\bibnamefont{Ries}},
\bibinfo{author}{\bibfnamefont{S.}~\bibnamefont{Roccia}},
\bibinfo{author}{\bibfnamefont{D.}~\bibnamefont{Rozpedzic}},
\bibinfo{author}{\bibfnamefont{A.}~\bibnamefont{Schnabel}},
\bibinfo{author}{\bibfnamefont{P.}~\bibnamefont{Schmidt-Wellenburg}},
\bibinfo{author}{\bibfnamefont{N.}~\bibnamefont{Severijns}},
\bibinfo{author}{\bibfnamefont{D.}~\bibnamefont{Shiers}},
\bibinfo{author}{\bibfnamefont{J.A.}~\bibnamefont{Thome}},
\bibinfo{author}{\bibfnamefont{A.}~\bibnamefont{Weis}},
\bibinfo{author}{\bibfnamefont{O.J.}~\bibnamefont{Winston}},
\bibinfo{author}{\bibfnamefont{E.}~\bibnamefont{Wursten}},
\bibinfo{author}{\bibfnamefont{J.}~\bibnamefont{ Zejma}} \bibnamefont{and}
\bibinfo{author}{\bibfnamefont{G.}~\bibnamefont{ Zsigmond}},
\bibinfo{title}{Revised experimental upper limit on the electric dipole moment of the neutron},
\bibinfo{journal}{Physical Review D} 
\textbf{\bibinfo{volume}{92}}, \bibinfo{pages}{092003}
  (\bibinfo{year}{2015}).



  
  


  


  
  











  
  
\bibitem[{\citenamefont{Abel, Afach, Ayres, Baker, Ban, Bison, Bodek, Bondar, Burghoff,  Chanel,  Chowdhuri, Chiu, Clement, Crawford, Daum, Emmenegger, Ferraris-Bouchez, Fertl, Flaux, Franke,  Fratangelo, Geltenbort,  Green, Griffith, van der Grinten, Grujić, Harris, Hayen, Heil, Henneck, Hélaine, Hild, Hodge, Horras, Iaydjiev, Ivanov, Kasprzak, Kermaidic, Kirch, Knecht, Knowles,  Koch,  Koss, Komposch, Kozela, Kraft, Krempel, Kuźniak, Lauss, Lefort, Lemière, Leredde, Mohanmurthy, Mtchedlishvili,  Musgrave, Naviliat-Cuncic, Pais, Piegsa, Pierre, Pignol, Plonka-Spehr, Prashanth, Quéméner, Rawlik, Rebreyend, Rienäcker, Ries, Roccia, Rogel, Rozpedzik, Schnabel, Schmidt-Wellenburg, Severijns, Shiers, Tavakoli Dinani, Thorne, Virot, Voigt, Weis, Wursten, Wyszynski, Zejma, Zenner and Zsigmond}(2020)}]{Afach2020}
\bibinfo{author}{\bibfnamefont{C.}~\bibnamefont{Abel}},
\bibinfo{author}{\bibfnamefont{S.}~\bibnamefont{Afach}},
\bibinfo{author}{\bibfnamefont{ N.J.}~\bibnamefont{Ayres}},
\bibinfo{author}{\bibfnamefont{ C.A}~\bibnamefont{Baker}},
\bibinfo{author}{\bibfnamefont{G.}~\bibnamefont{Ban}},
\bibinfo{author}{\bibfnamefont{G.}~\bibnamefont{Bison}},
\bibinfo{author}{\bibfnamefont{K.}~\bibnamefont{Bodek}},
\bibinfo{author}{\bibfnamefont{V.}~\bibnamefont{Bondar}},
\bibinfo{author}{\bibfnamefont{M.}~\bibnamefont{Burghoff}},
\bibinfo{author}{\bibfnamefont{E.}~\bibnamefont{Chanel}},
\bibinfo{author}{\bibfnamefont{Z.}~\bibnamefont{Chowdhuri}},
\bibinfo{author}{\bibfnamefont{P.-J.}~\bibnamefont{Chiu}},
\bibinfo{author}{\bibfnamefont{B.}~\bibnamefont{Clement}},
\bibinfo{author}{\bibfnamefont{C.B.}~\bibnamefont{Crawford}},
\bibinfo{author}{\bibfnamefont{M.}~\bibnamefont{Daum}},
\bibinfo{author}{\bibfnamefont{S.}~\bibnamefont{Emmenegger}},
\bibinfo{author}{\bibfnamefont{L.}~\bibnamefont{Ferraris-Bouches}},
\bibinfo{author}{\bibfnamefont{M.}~\bibnamefont{Fertl}},
\bibinfo{author}{\bibfnamefont{P.}~\bibnamefont{Flaux}},
\bibinfo{author}{\bibfnamefont{B.}~\bibnamefont{ Franke}},
\bibinfo{author}{\bibfnamefont{A.}~\bibnamefont{Fratangelo}},
\bibinfo{author}{\bibfnamefont{P.}~\bibnamefont{Geltenbort}},
\bibinfo{author}{\bibfnamefont{K.}~\bibnamefont{Green}},
\bibinfo{author}{\bibfnamefont{W.C.}~\bibnamefont{Griffith}},
\bibinfo{author}{\bibfnamefont{M.}~\bibnamefont{van der Grinten}},
\bibinfo{author}{\bibfnamefont{Z.D.}~\bibnamefont{Grujić}},
\bibinfo{author}{\bibfnamefont{P.G.}~\bibnamefont{Harris}},
\bibinfo{author}{\bibfnamefont{L.}~\bibnamefont{Hayen}},
\bibinfo{author}{\bibfnamefont{W.}~\bibnamefont{ Heil}},
\bibinfo{author}{\bibfnamefont{R.}~\bibnamefont{Henneck}},
\bibinfo{author}{\bibfnamefont{V.}~\bibnamefont{ Hélaine}},
\bibinfo{author}{\bibfnamefont{N.}~\bibnamefont{Hild}},
\bibinfo{author}{\bibfnamefont{Z.}~\bibnamefont{Hodge}},
\bibinfo{author}{\bibfnamefont{M.}~\bibnamefont{ Horras}},
\bibinfo{author}{\bibfnamefont{P.}~\bibnamefont{Iaydjiev}},
\bibinfo{author}{\bibfnamefont{S.N.}~\bibnamefont{Ivanov}},
\bibinfo{author}{\bibfnamefont{M.}~\bibnamefont{Kasprzak}},
\bibinfo{author}{\bibfnamefont{Y.}~\bibnamefont{Kermaidic}},
\bibinfo{author}{\bibfnamefont{K.}~\bibnamefont{Kirch}},
\bibinfo{author}{\bibfnamefont{A.}~\bibnamefont{Knecht}},
\bibinfo{author}{\bibfnamefont{P.}~\bibnamefont{ Knowles}},
\bibinfo{author}{\bibfnamefont{H.-C.}~\bibnamefont{Koch}},
\bibinfo{author}{\bibfnamefont{P.A.}~\bibnamefont{Koss}},
\bibinfo{author}{\bibfnamefont{S.}~\bibnamefont{Komposch}},
\bibinfo{author}{\bibfnamefont{A.}~\bibnamefont{Kozela}},
\bibinfo{author}{\bibfnamefont{A.}~\bibnamefont{Kraft}},
\bibinfo{author}{\bibfnamefont{J.}~\bibnamefont{Krempel}},
\bibinfo{author}{\bibfnamefont{M.}~\bibnamefont{Kuźniak}},
\bibinfo{author}{\bibfnamefont{B.}~\bibnamefont{Lauss}},
\bibinfo{author}{\bibfnamefont{T.}~\bibnamefont{Lefort}},
\bibinfo{author}{\bibfnamefont{Y.}~\bibnamefont{ Lemiere}},
\bibinfo{author}{\bibfnamefont{A.}~\bibnamefont{Leredde}},
\bibinfo{author}{\bibfnamefont{P.}~\bibnamefont{Mohanmurthy}},
\bibinfo{author}{\bibfnamefont{A.}~\bibnamefont{Mtchedlishvili}},
\bibinfo{author}{\bibfnamefont{M.}~\bibnamefont{ Musgrave}},
\bibinfo{author}{\bibfnamefont{O.}~\bibnamefont{Naviliat-Cuncic}},
\bibinfo{author}{\bibfnamefont{D.}~\bibnamefont{Pais}},
\bibinfo{author}{\bibfnamefont{F.M.}~\bibnamefont{Piegsa}},
\bibinfo{author}{\bibfnamefont{E.}~\bibnamefont{Pierre}},
\bibinfo{author}{\bibfnamefont{G.}~\bibnamefont{Pignol}},
\bibinfo{author}{\bibfnamefont{C.}~\bibnamefont{Plonka-Spehr}},
\bibinfo{author}{\bibfnamefont{P.N.}~\bibnamefont{ Prashanth}},
\bibinfo{author}{\bibfnamefont{G.}~\bibnamefont{Quemener}},
\bibinfo{author}{\bibfnamefont{M.}~\bibnamefont{Rawlik}},
\bibinfo{author}{\bibfnamefont{D.}~\bibnamefont{Rebreyend}},
\bibinfo{author}{\bibfnamefont{I.}~\bibnamefont{ Rienäcker}},
\bibinfo{author}{\bibfnamefont{D.}~\bibnamefont{Ries}},
\bibinfo{author}{\bibfnamefont{S.}~\bibnamefont{Roccia}},
\bibinfo{author}{\bibfnamefont{G.}~\bibnamefont{Rogel}},
\bibinfo{author}{\bibfnamefont{D.}~\bibnamefont{Rozpedzic}},
\bibinfo{author}{\bibfnamefont{A.}~\bibnamefont{Schnabel}},
\bibinfo{author}{\bibfnamefont{P.}~\bibnamefont{Schmidt-Wellenburg}},
\bibinfo{author}{\bibfnamefont{N.}~\bibnamefont{Severijns}},
\bibinfo{author}{\bibfnamefont{D.}~\bibnamefont{Shiers}},
\bibinfo{author}{\bibfnamefont{R.}~\bibnamefont{Tavakoli Dinani}},
\bibinfo{author}{\bibfnamefont{J.A.}~\bibnamefont{Thorne}},
\bibinfo{author}{\bibfnamefont{R.}~\bibnamefont{Virot}},
\bibinfo{author}{\bibfnamefont{J.}~\bibnamefont{Voigt}},
\bibinfo{author}{\bibfnamefont{A.}~\bibnamefont{Weis}},
\bibinfo{author}{\bibfnamefont{E.}~\bibnamefont{Wursten}},
\bibinfo{author}{\bibfnamefont{G.}~\bibnamefont{ Wyszynski}},
\bibinfo{author}{\bibfnamefont{J.}~\bibnamefont{ Zejma}},
\bibinfo{author}{\bibfnamefont{J.}~\bibnamefont{Zenner}} \bibnamefont{and}
\bibinfo{author}{\bibfnamefont{G.}~\bibnamefont{Zsigmond}},
\bibinfo{title}{Measurement of the permanent electric dipole moment of the neutron},
\bibinfo{journal}{Physical Review Letters} 
\textbf{\bibinfo{volume}{124}}, \bibinfo{pages}{081803}
(\bibinfo{year}{2020}).





\bibitem[{\citenamefont{Landau and Lifshitz}(1971)}]{Landau2}
\bibinfo{author}{\bibfnamefont{L.D.}~\bibnamefont{Landau}}  \bibnamefont{and}
\bibinfo{author}{\bibfnamefont{E.M.}~\bibnamefont{Lifshitz}},
\bibinfo{title}{The Classical Theory of Fields},
  \bibinfo{journal}{Pergamon Press. Oxford - New-York - Toronto - Sydney - Braunschweig} \textbf{\bibinfo{volume}{2}},
  \bibinfo{pages}{} (\bibinfo{year}{1971}).
  



















\end{thebibliography}
\end{document}